\documentclass[twocolumn]{aastex631}
\usepackage{graphicx}
\usepackage[caption=false]{subfig}
\usepackage{amsmath}
\usepackage{booktabs}
\usepackage[T1]{fontenc}
\usepackage{xcolor}
\usepackage{censor}

\let\pwiflocal=\iffalse \let\pwifjournal=\iffalse

\newcommand{\hatp}{\object{HAT-P-67}~}
\newcommand{\hatpb}{\object{HAT-P-67 b}}

\newcommand{\ms}{\ensuremath{\rm m\,s^{-1}}}

\newcommand{\rsun}{\ensuremath{R_\sun}}

\newcommand{\rstar}{\ensuremath{R_\star}}

\newcommand{\rpl}{\ensuremath{R_{p}}}
\newcommand{\mpl}{\ensuremath{M_{p}}}
\newcommand{\kms}{\ensuremath{\rm km\,s^{-1}}}

\newcommand{\arstar}{\ensuremath{a/\rstar}}

\newcommand{\mjup}{\ensuremath{M_{\rm J}}}

\providecommand{\adsurl}[1]{\href{#1}{ADS}}

\newcommand{\PSUAA}{Department of Astronomy \& Astrophysics, 525 Davey Laboratory, The Pennsylvania State University, University Park, PA, 16802, USA}
\newcommand{\PSUCEHW}{Center for Exoplanets and Habitable Worlds, 525 Davey Laboratory, The Pennsylvania State University, University Park, PA, 16802, USA}
\newcommand{\Princeton}{Department of Astrophysical Sciences, Princeton University, 4 Ivy Lane, Princeton, NJ 08540, USA}

\newcommand{\CFA}{Center for Astrophysics $\vert$ Harvard $\&$ Smithsonian, 60 Garden Street, MS-16, Cambridge, MA 02138, USA}
\newcommand{\UTAustin}{Department of Astronomy, The University of Texas at Austin, 2515 Speedway, Austin, TX 78712, USA}
\newcommand{\MCDonald}{McDonald Observatory and Department of Astronomy, The University of Texas at Austin, 2515 Speedway, Austin, TX 78712, USA}
\newcommand{\UTSpace}{Center for Planetary Systems Habitability, The University of Texas at Austin, 2515 Speedway, Austin, TX 78712, USA}
\newcommand{\Amsterdam}{Anton Pannekoek Institute for Astronomy, University of Amsterdam, Science Park 904, NL-1098 XH Amsterdam, The Netherlands}
\newcommand{\UCSC}{Department of Astronomy \& Astrophysics, University of California, Santa Cruz, 1156 High St, Santa Cruz, CA 95064, USA}
\newcommand{\Tata}{Department of Astronomy and Astrophysics, Tata Institute of Fundamental Research, Homi Bhabha Road, Colaba, Mumbai 400005, India}
\newcommand{\Steward}{Steward Observatory, The University of Arizona, 933 N. Cherry Ave, Tucson, AZ 85721, USA}
\newcommand{\HET}{Hobby-Eberly Telescope, The University of Texas at Austin, 2515 Speedway, Austin, TX 78712, USA}
\newcommand{\MIT}{Department of Physics and Kavli Institute for Astrophysics and Space Research, Massachusetts Institute of Technology, Cambridge, MA 02139, USA}

\begin{document}
\shorttitle{Runaway Mass Loss in the HAT-P-67 system}
\shortauthors{Gully-Santiago et al.}

\title{A Large and Variable Leading Tail of Helium in a Hot Saturn Undergoing Runaway Inflation}

\author[0000-0002-4020-3457]{Michael Gully-Santiago}
\affil{\UTAustin}

\author[0000-0002-4404-0456]{Caroline V. Morley}
\affil{\UTAustin}

\author[0000-0003-2152-9248]{Jessica Luna}
\affil{\UTAustin}

\author[0000-0002-1417-8024]{Morgan MacLeod}
\affiliation{\CFA}

\author[0000-0002-9584-6476]{Antonija Oklop{\v{c}}i{\'c}}
\affil{\Amsterdam}

\author[0000-0002-1846-196X]{Aishwarya Ganesh}
\affil{\UTAustin}

\author[0000-0001-6532-6755]{Quang H. Tran}
\affiliation{\UTAustin}

\author[0000-0002-3726-4881]{Zhoujian Zhang}\thanks{NASA Sagan Fellows}
\affiliation{\UCSC}

\author[0000-0003-2649-2288]{Brendan P. Bowler}
\affiliation{\UTAustin}

\author[0000-0001-9662-3496]{William D. Cochran}
\affil{\MCDonald}
\affil{\UTSpace}

\author[0000-0001-9626-0613]{Daniel M. Krolikowski}
\affil{\Steward}

\author[0000-0001-9596-7983]{Suvrath Mahadevan}
\affil{\PSUAA}
\affil{\PSUCEHW}

\author[0000-0001-8720-5612]{Joe P.\ Ninan}
\affil{\Tata}

\author[0000-0001-7409-5688]{Guðmundur Stefánsson}\thanks{NASA Sagan Fellows}
\affil{\Princeton}

\author[0000-0001-7246-5438]{Andrew Vanderburg}
\affil{\MIT}

\author[0000-0002-2259-4116]{Joseph A. Zalesky}
\affil{\UTAustin}

\author[0000-0003-2307-0629]{Gregory R. Zeimann}
\affil{\HET}


\begin{abstract}
    Atmospheric escape shapes the fate of exoplanets, with statistical evidence for transformative mass loss imprinted across the mass-radius-insolation distribution. Here we present transit spectroscopy of the highly irradiated, low-gravity, inflated hot Saturn HAT-P-67~b.  The Habitable Zone Planet Finder (HPF) spectra show a detection of up to 10\% absorption depth of the 10833 \AA~ Helium triplet. The 13.8 hours of on-sky integration time over 39 nights sample the entire planet orbit, uncovering excess Helium absorption preceding the transit by up to 130 planetary radii in a large leading tail. This configuration can be understood as the escaping material overflowing its small Roche lobe and advecting most of the gas into the stellar---and not planetary---rest frame, consistent with the Doppler velocity structure seen in the Helium line profiles. The prominent leading tail serves as direct evidence for dayside mass loss with a strong day-/night- side asymmetry. We see some transit-to-transit variability in the line profile, consistent with the interplay of stellar and planetary winds. We employ 1D Parker wind models to estimate the mass loss rate, finding values on the order of $2\times10^{13}$ g/s, with large uncertainties owing to the unknown XUV flux of the F host star. The large mass loss in HAT-P-67 b represents a valuable example of an inflated hot Saturn, a class of planets recently identified to be rare as their atmospheres are predicted to evaporate quickly. We contrast two physical mechanisms for runaway evaporation: Ohmic dissipation and XUV irradiation, slightly favoring the latter.   
\end{abstract}


\section{Introduction}\label{sec:intro}

Atmospheric escape appears to be a normal phase for exoplanets of a certain size and insolation, statistically imprinted in the dearth of 1.5$-$2.0~$R_\oplus$ planets within 100-day periods \citep{2017AJ....154..109F}.  The leading candidate physical mechanisms for this rapid transition between mini-Neptunes and super-Earths include photoevaporation \citep{2013ApJ...775..105O,2013ApJ...776....2L,2017ApJ...847...29O} and core-powered mass loss \citep{2019MNRAS.487...24G,2023arXiv230200009B}.  Whatever the cause, some large fraction of planets undergo transformative atmospheric escape, and the signal should be widely discernable.  Such signals have been searched for, and increasingly found, in many transiting planet systems with at least 28 detections to date \citep{2022arXiv221116243D}.

Uncertainty in system ages, evaporation timescales, X-ray/UV radiation, and dominating physical mechanisms degrade our ability to foretell if any given planet will exhibit ongoing signatures of atmospheric escape.  Episodic stellar wind gusts and other forms of astrophysical variability could also subdue the appearance of atmospheric escape, even where we expect it most.  The over 57 published non-detections of atmospheric escape \citep{2022arXiv221116243D,2023arXiv230700967G} must encode these natural whims in a way that we have not yet disentangled. Nevertheless, we can boost our chances of witnessing active and significant atmospheric escape by targeting sources that seem predisposed to loss.  These intrinsic or extrinsic factors may include proximity to the host star, low surface gravity, and high energy incident radiation.

Inflated hot Saturns stand out as an especially extreme category that should exhibit mass loss.  This category is defined as having masses comparable to Saturn's ($M_p \sim 0.3 M_\mathrm{Jup}$), with equilibrium temperatures high enough to expect radius inflation ($T_\mathrm{eq}>1000\;$K).  Their low gravitational potentials should let go of their atmospheres more readily than their hot Jupiter counterparts.  Lower gravity also implies larger atmospheric scale heights, making them easier to detect in transmission spectroscopy.  And their large transit depths and short periods should make them readily detectable in transit searches in large numbers, like hot Jupiters.

But inflated hot Saturns are rare \citep{2018AJ....155..214T}.  The cause for their underabundance remains an open question, with at least two conceivable explanations.  Tidal migration mechanisms---either high-eccentricity or disk-based---could hypothetically proceed in a mass-dependent manner, efficiently for Jupiter-mass planets, but inefficiently for the lower-mass Saturns \citep{2018AJ....155..214T,2018ARA&A..56..175D}.  In this scenario, sub-Saturn mass planets never make it to the close-in orbital separations that would lead to the conditions needed for inflation in the first place.

Alternatively---and most consequentially for atmospheric escape---another explanation may prevail.  Inflated hot Saturns may either form \emph{in-situ} or effectively migrate to close-in orbital separations \citep{2018ARA&A..56..175D}, but once they arrive, the intense irradiation overheats the planet.  This heating leads to runaway inflation and, ultimately, complete disintegration.  In this scenario, the inflationary half-life becomes so short that the probability of observing members in the class decreases sharply with density, causing the apparent lack of inflated sub-Saturns \citep{2023ApJ...945L..36T}.

\citet{2011ApJ...738....1B} predicted runaway inflation of hot Saturns as a consequence of the Ohmic dissipation mechanism.  Here, lightly thermally ionized atmospheric flows induce drag in a planetary magnetic field, weakly coupling the stellar incident energy into the planetary interior.  A key prediction of Ohmic dissipation is that the anomalous heating efficiency $\epsilon(T_\mathrm{eq})$ should exhibit a peak around $T_\mathrm{eq}\sim1500-2000\;$K \citep{2012ApJ...745..138M,2014ApJ...794..132R,2016ApJ...819..116G}. \citet{2018AJ....155..214T} favored Ohmic dissipation as the mechanism responsible for inflating hot Jupiters by showing that the observed sample of inflated planets implies an anomalous heating efficiency peak at equilibrium temperatures of $\sim1500\;$K.  Therefore Ohmic dissipation stands out as a leading physical mechanism for inflation and mass loss.

Recently, \citet{2023ApJ...945L..36T} showed that hot Saturns can undergo catastrophic erosion by stellar X-ray and Extreme UV (XUV) photoevaporative mass loss. Planets with densities less than $\sim$0.1 g~cm$^{-3}$ achieve mass loss rates of up to $10^3$ $M_\oplus /$Gyr, setting up a positive feedback loop: the planet increases in radius, overflowing its Roche lobe, fueling greater mass loss, and increasing in radius even further.  This vicious cycle systematically depopulates the mass-radius plane with a cliff defined by the $\rho_p \sim$0.1 g~cm$^{-3}$ dividing line.

A runaway inflation scenario predicts an inevitable and profound mass loss rate for inflated hot Saturns.  As the planet's atmosphere overflows its Roche lobe at an ever-increasing pace, instantaneous mass loss rates may exceed $\dot{M}>10^{13}$ g/s, over 10$\times$ larger than those previously seen in atmospheric escape measurements to date \citep{2022arXiv221116243D}.  Inflated hot Saturns, therefore, make excellent targets for direct measurement of atmospheric escape and for testing the underlying mechanisms of mass loss.

Large mass loss alone does not guarantee detectability.  The ability to detect even immense mass loss hinges on its observability in spectral tracers.  Ly$\alpha$, \ion{He}{1} 10833 \AA, and H$\alpha$ have emerged as the most amenable to detection \citep{2000ApJ...537..916S,2003Natur.422..143V,2012ApJ...751...86J,2018NatAs...2..714Y,2018ApJ...855L..11O,2018Natur.557...68S, 2022arXiv221116243D,2023MNRAS.518.4357O}, but each of these has its own observational limitations.  Ly$\alpha$ can suffer from interstellar medium \ion{H}{1} absorption censoring its low-velocity line core, for example.  Here we focus on \ion{He}{1} 10833 \AA, which offers some advantages.  In particular, metastable Helium's ability to be observed at high spectral resolution from the ground has been especially valuable for evincing velocity substructure of the escaping gas motion relative to the exoplanet restframe \citep{2019A&A...629A.110A,2020ApJ...894...97N}, and convincingly associating the signal to an exoplanetary origin as opposed to stellar contamination \citep{2018AJ....156..189C}.  \ion{He}{1} has resulted in at least 14 systems with detections \citep{2022arXiv221116243D}\footnote{See also Table~S1 of \citealt{doi:10.1126/sciadv.adf8736} for a compilation of exoplanets with detections and non-detections of the helium excess.}.  The sample of detections includes both hot Jupiters and lower mass planets but lacks inflated hot Saturns owing to their intrinsic rarity below the 0.1 g~cm$^{-3}$ threshold. An observational picture of mass loss in inflated hot Saturns appears to be lacking for this reason.

\begin{figure}
    \centering
    \includegraphics[width=\linewidth]{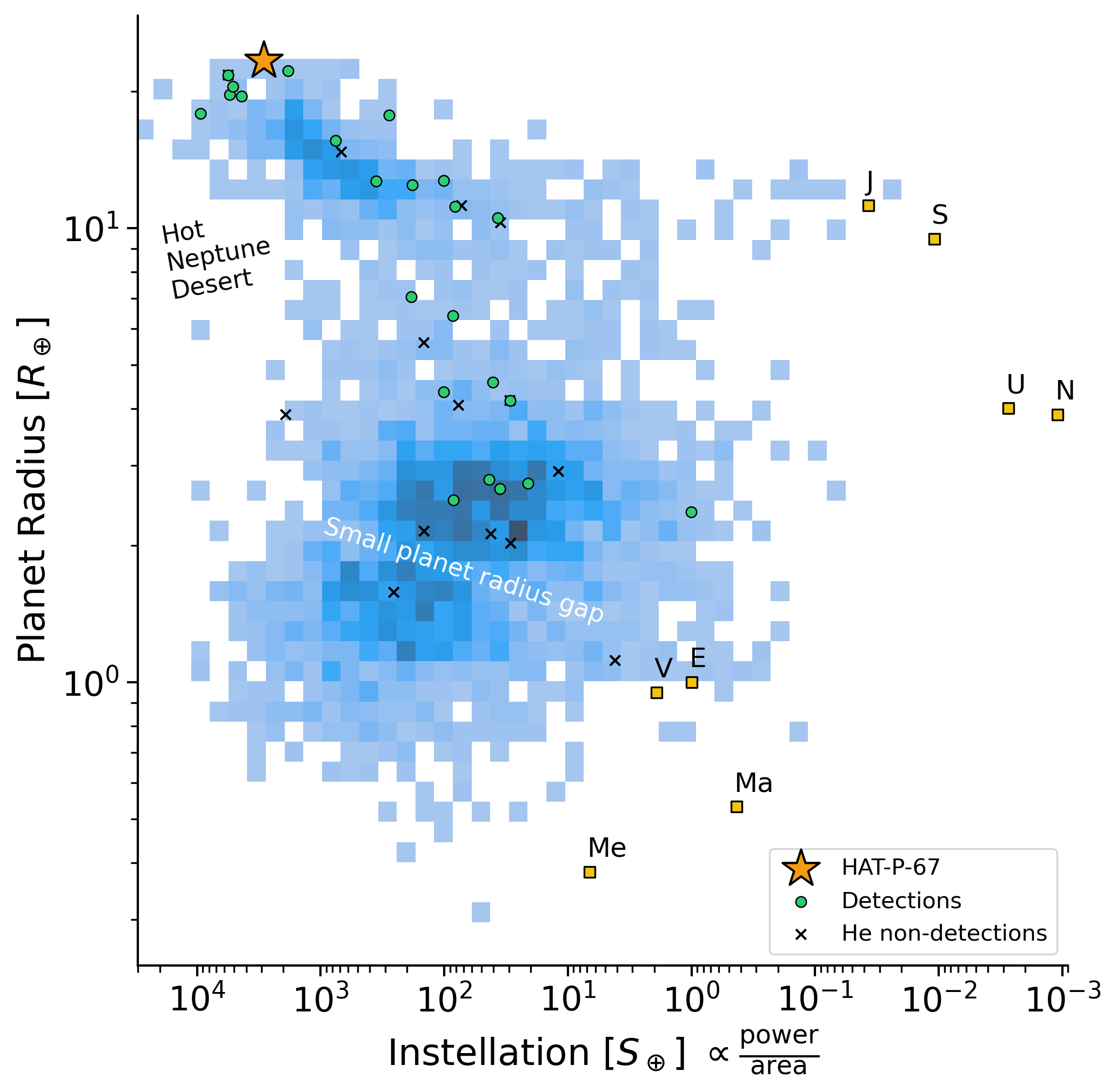}
    \caption{Overview of exoplanet demographics.  The pixel bins reflect the observed density of over 5000 planets accessed from the NASA Exoplanet Archive. Detections of atmospheric escape are common among large planets with strong insolation.  \hatpb is among the most inflated planets known.}
    \label{fig:instellation}
\end{figure}

\begin{figure}
    \includegraphics[width=\linewidth]{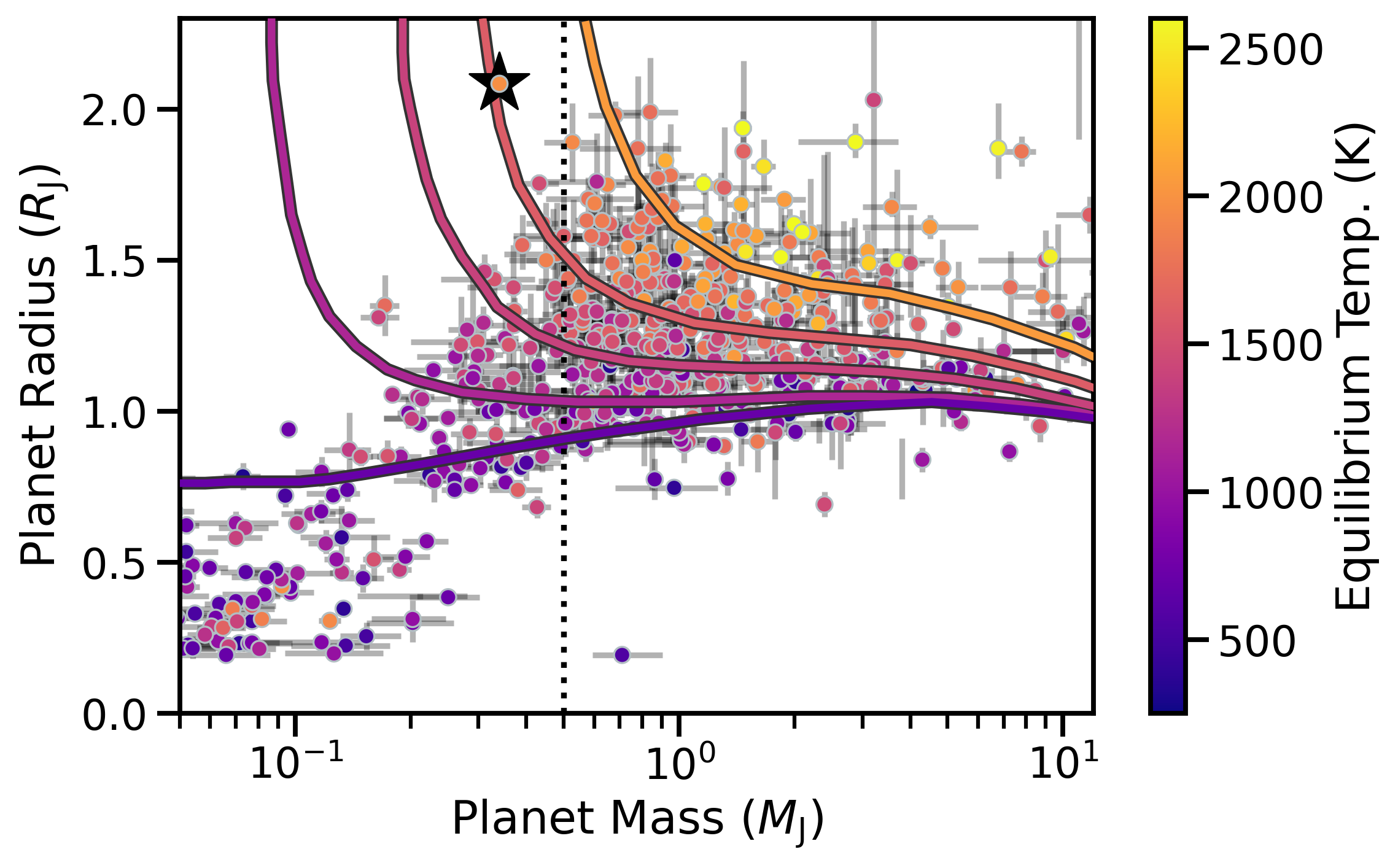}
    \caption{Mass-radius trends for inflated hot sub-Saturns and hot Jupiters, with layout following Figure 2 of \citet{2018AJ....155..214T} and updated with NASA Exoplanet Archive confirmed planets.  The trend lines show the mass-radius relationship for equilibrium temperatures of 500, 1000, 1250, 1500, and 2000 K, assuming the mean composition and mean heating efficiency from the original figure.  HAT-P-67~b ($\star$ symbol) stands alone in a region defined by the lack of inflated sub-Saturns and explained by short lifetimes from runaway inflation and mass loss.}
    \label{fig:ThornFortUpdated}
\end{figure}

The Habitable Zone Planet Finder (HPF) Helium Exospheres program has been conducting a survey of exoplanets to search for atmosphere loss via the \ion{He}{1} 10833 \AA~ metastable triplet.  The survey's multiple-visit sampling strategy has enabled a search for atmospheric loss at large out-of-transit separations from the planet, recently revealing giant tidal tails of Helium escaping the hot Jupiter HAT-P-32~b \citep{doi:10.1126/sciadv.adf8736}. 

Here we present a multi-year observational campaign searching for atmospheric escape in \object{HAT-P-67 b}, a transiting inflated hot Saturn orbiting an F5 subgiant at an orbital separation of 0.06 AU and a 4.8-day orbital period \citep{2017AJ....153..211Z}.  Its strong insolation ($T_\mathrm{eq}\sim2000\;$K) combined with \object{HAT-P-67 b}'s extremely low surface gravity ($\log{g_p}<2.3$~dex) makes it an exceptional candidate for strong atmospheric mass loss via Roche lobe overflow.  Its $\sim$0.05 g~cm$^{-3}$ density places it below the 0.1 g~cm$^{-3}$ threshold predicted to exhibit runaway inflation \citep{2023ApJ...945L..36T}.  Figures \ref{fig:instellation} and \ref{fig:ThornFortUpdated} show how much of an outlier \hatpb~ is: large, low mass, and heavily irradiated.    

The evolutionary state of HAT-P-67~b offers even more intriguing possibilities.  The planet may have undergone re-inflation \citep{2022AJ....163...53S, 2022AJ....163..120G, 2023arXiv230306728G}, as the host star evolved through the subgiant phase-- with the planet's insolation increasing with the star's rapid luminosity jump in this part of the short-lived HR diagram.  The tidal gravity of the nearby massive ($M_\star \sim1.6\;M_\odot$) host star may amplify mass loss rates even further \citep{2007A&A...472..329E, 2023ApJ...945L..36T}. Its status as a rare inflated hot Saturn makes HAT-P-67~b a promising laboratory, uniquely suited for testing the runaway inflationary predictions of Ohmic dissipation and XUV photoevaporation.

We assemble both archival, previously published, and new observations of the HAT-P-67 system, chronicled in Section \ref{secObs}.  In Section \ref{secAnalysis}, we refine the stellar and planet properties based on those observations, including distance, radius, and orbit re-analyses.  The Helium excess detection is presented in Section \ref{secHeAnalysis}, with an analysis of the signal's trend with orbital phase.  We use this orbital structure to reconstruct the geometry and mass loss of the escaping Helium exosphere with 1D Parker wind models (Section \ref{secResults}).  We assess the physical mechanisms (Section \ref{secPhysMech}) giving rise to the escaping atmosphere, weighing the distinctive predictions of XUV irradiation and Ohmic dissipation.  Finally, in Section \ref{secDiscuss}, we question the assumptions in our approach, discuss the overall congruence of predictions and observations, and highlight some implications for future exosphere studies.

\section{Observations}\label{secObs}
\subsection{Habitable Zone Planet Finder (HPF)}

The Habitable Zone Planet Finder Spectrograph \citep[HPF;][]{2012SPIE.8446E..1SM,2014SPIE.9147E..1GM, 2019Optic...6..233M} on the queue-scheduled 10-meter \emph{Hobby-Eberly Telescope} \citep[HET;][]{1998SPIE.3352...34R} operates in the near-IR from $8100-12800~$\AA~ spanning the \textit{z}, \textit{Y}, and \textit{J} bands at spectral resolving power $R=55,000$. The HET fixed-elevation design \citep{2007PASP..119..556S} limits the observability of \object{HAT-P-67} to less than 1 hour ``tracks'' for a fixed range of hour angles before (east track) and after (west track) the star transits the meridian.  Whereas conventional steerable telescopes could conduct continuous point-and-stare observations of \object{HAT-P-67} for hours, HET cannot.  In practice, this limitation means that in-transit and out-of-transit observational phases were rarely possible on the same night.  Instead, we organized the observations into four campaigns to coincide with \object{HAT-P-67 b} transits on UT dates 2020 April 28, 2020; May 22, 2020; June 15, 2020; and 2022 April 29.  These campaigns have out-of-transit observations at least one night before and one night after, and often two nights on either side of the transit.  Two more transit snapshots were obtained in 2022 June-July without the accompanying visits immediately before and after.  The in-transit campaigns had up to 14 exposures per HET track, with integration times between 5-8.5 minutes.  We also obtained random-in-phase lower-priority ``P1-P4'' reconnaissance observations \citep{2007PASP..119..556S}.  These out-of-transit snapshot observations typically received 4 or fewer individual exposures.  We observed HAT-P-67 with HPF for a total of 41 visits on 39 unique nights, with two of those nights observing both the east and west HET tracks.  The total on-source integration time exceeds 13.8 hours, with $N=152$ individual spectra possessing a typical signal-to-noise ratio of 80 per pixel.  Table \ref{tabHPFLog} summarizes the log of observations.

The observation schedule was strategically restricted to the spring season when the Barycentric Earth Radial Velocity (BERV) would Doppler shift telluric absorption lines sufficiently far away from the core of the Helium 10833 \AA~ feature \citep{2022ApJ...939L..11S}.  The small BERV still means the redward Helium line wing has significant telluric contamination between 10834$-$10836 \AA, but the core and blue line wings appear relatively pristine.  Sky emission lines land in this region but are more easily mitigated by HPF's simultaneous sky reference fiber.

\begin{figure*}
    \centering
    \includegraphics[width=0.98\linewidth]{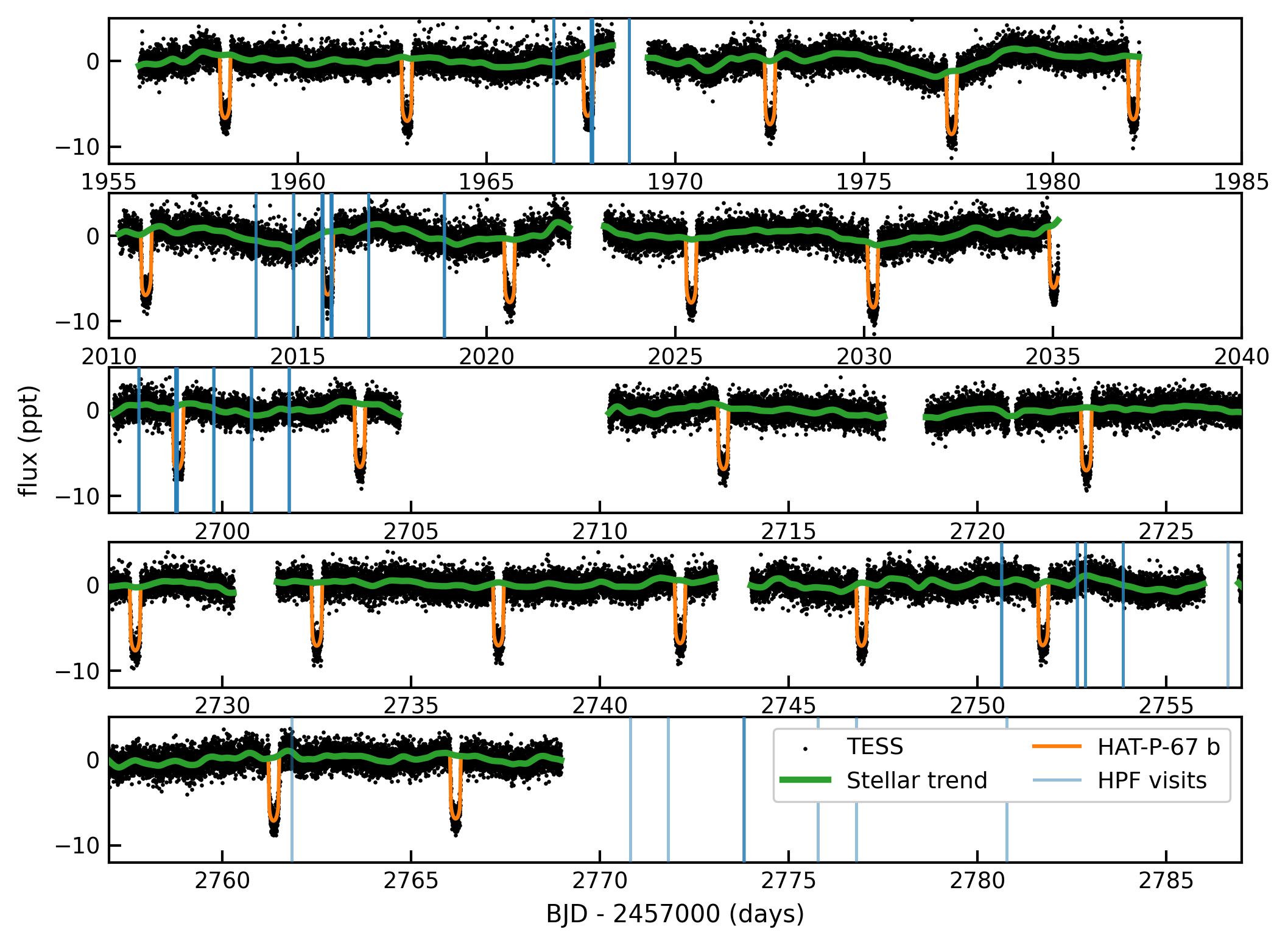}
    \caption{Overview of all available TESS Sectors showing 24 full or partial transits with 34 visits with HPF (vertical gray bars), 7 of which coincide with transits.  The rest of the visits sample out-of-transit phases.  The 8 HPF visits between days 2050 and 2690 are not shown.  Note the large time breaks between some consecutive panels, which correspond roughly to the duration of a TESS sector.}
    \label{fig:TESSoverview}
\end{figure*}

Only 19 out of the 39 nights possess A0V telluric calibration standard stars.  These standard stars were used to spot-check our telluric masking.  Figure \ref{fig:TESSoverview} indicates the epochs of select HPF visits overlaid on the \emph{TESS} time-series lightcurve.

\begin{deluxetable}{lccrchr}
    \tablewidth{0pc}
    \tabletypesize{\scriptsize}
    \tablecaption{
        HPF Observation Log
        \label{tabHPFLog}
    }
    \tablehead{
        \colhead{UTC}   &
        \colhead{Track} &
        \colhead{Desc.} &
        \colhead{BJD} &
        \colhead{$N_\mathrm{exp}$} &
        \nocolhead{$t_\mathrm{exp}$} &
        \colhead{$\phi$}\\
        \colhead{}   &
        \colhead{} &
        \colhead{} &
        \colhead{2457000.00+} &
        \colhead{} &
        \nocolhead{(s)} &
        \colhead{}
    }
    \startdata
    2020-04-27 & E & Pre & 1966.78 & 4 & 308.85 & -0.192 \\
2020-04-28 & E & Transit & 1967.79 & 12 & 308.85 & 0.018 \\
2020-04-29 & E & Control & 1968.78 & 4 & 308.85 & 0.225 \\
2020-05-20 & W & Pre & 1989.95 & 4 & 308.85 & -0.373 \\
2020-05-21 & W & Pre & 1990.94 & 4 & 308.85 & -0.169 \\
2020-05-22 & E & Transit & 1991.72 & 14 & 308.85 & -0.006 \\
2020-05-23 & W & Control & 1992.93 & 4 & 308.85 & 0.246 \\
2020-05-24 & W & Baseline & 1993.94 & 4 & 308.85 & 0.455 \\
2020-06-13 & W & Pre & 2013.89 & 3 & 308.85 & -0.397 \\
2020-06-14 & W & Pre & 2014.89 & 4 & 308.85 & -0.190 \\
2020-06-15 & E & Pre & 2015.64 & 5 & 308.85 & -0.033 \\
2020-06-15 & E & Transit & 2015.66 & 4 & 308.85 & -0.029 \\
2020-06-15 & W & Transit & 2015.89 & 9 & 308.85 & 0.019 \\
2020-06-16 & W & Control & 2016.87 & 4 & 308.85 & 0.224 \\
2020-06-18 & W & Pre & 2018.88 & 4 & 308.85 & -0.360 \\
2020-07-22 & W & Pre & 2052.79 & 2 & 511.20 & -0.309 \\
2020-08-01 & W & Pre & 2062.75 & 2 & 511.20 & -0.240 \\
2021-01-31 & E & Pre & 2246.03 & 2 & 511.20 & -0.136 \\
2021-02-01 & E & Post & 2247.02 & 2 & 511.20 & 0.070 \\
2021-02-24 & E & Pre & 2269.96 & 1 & 511.20 & -0.161 \\
2021-02-26 & E & Control & 2271.94 & 2 & 511.20 & 0.250 \\
2021-03-04 & E & Baseline & 2277.93 & 2 & 511.20 & 0.496 \\
2021-03-31 & E & Post & 2304.86 & 2 & 511.20 & 0.094 \\
2022-04-28 & E & Pre & 2697.79 & 4 & 308.85 & -0.217 \\
2022-04-29 & E & Transit & 2698.78 & 14 & 308.85 & -0.011 \\
2022-04-30 & E & Control & 2699.77 & 4 & 308.85 & 0.196 \\
2022-05-01 & E & Baseline & 2700.77 & 4 & 308.85 & 0.403 \\
2022-05-02 & E & Pre & 2701.77 & 4 & 308.85 & -0.390 \\
2022-06-20 & E & Pre & 2750.64 & 3 & 308.85 & -0.230 \\
2022-06-22 & E & Control & 2752.64 & 3 & 308.85 & 0.187 \\
2022-06-22 & W & Control & 2752.86 & 3 & 308.85 & 0.232 \\
2022-06-23 & W & Baseline & 2753.86 & 3 & 308.85 & 0.439 \\
2022-06-26 & E & Transit & 2756.64 & 1 & 308.85 & 0.017 \\
2022-07-01 & W & Post & 2761.85 & 1 & 308.85 & 0.101 \\
2022-07-10 & W & Pre & 2770.82 & 1 & 308.85 & -0.035 \\
2022-07-11 & W & Control & 2771.81 & 1 & 308.85 & 0.172 \\
2022-07-13 & W & Pre & 2773.82 & 3 & 308.85 & -0.411 \\
2022-07-15 & W & Transit & 2775.78 & 1 & 308.85 & -0.003 \\
2022-07-16 & W & Control & 2776.80 & 1 & 308.85 & 0.209 \\
2022-07-20 & W & Post & 2780.77 & 1 & 308.85 & 0.034 \\
2022-07-29 & W & Pre & 2789.74 & 1 & 308.85 & -0.101 \\
2022-07-30 & W & Post & 2790.78 & 1 & 308.85 & 0.115 \\
    \enddata
    \tablecomments{All exposure times were 308.85~s except for observations from 2020-07-22 to 2021-03-31, which were 511.20~s.  Column 2 describes the HET ``track'', restricted to either East (E) or West (W).  Column 5 is the normalized phase $\phi \in(-0.5, 0.5)$, with zero at midtransit.}
\end{deluxetable}

\subsection{TESS Light Curves}
HAT-P-67 was observed with the \emph{Transiting Exoplanet Survey Satellite} \citep[TESS,][]{2014SPIE.9143E..20R} in Sectors 24, 26, 51, 52, 53 with 2-minute cadence, and in Sector 25 with 30-minute (FFI) cadence.  The Sector 25 FFI data fell just barely off the TESS detectors, in collateral pixels where no starlight ever hit the detector.

We also assembled a comparison sample of about 1000 lightcurves to interpret the prevalence of lightcurve modulation and stellar activity among broadly F subgiant-like stars.  We selected sources based on \emph{Gaia} DR3 $T_\mathrm{eff}$ estimates and similar $\log{g}$,  and availability of at least one sector of TESS 2-minute cadence data.  We visually spot-checked these lightcurves to understand artifacts and windowing effects.

\subsection{Gaia DR3}\label{gaiadr3}
The stellar system consists of a binary with an M dwarf companion HAT-P-67B (\emph{Gaia DR3 1358614983131339904}) separated on-sky by $9\farcs0$ \citep{2019MNRAS.490.5088M}, well-separated from the planet-host star and not a source of contamination for the HPF observations.  HAT-P-67A (\emph{Gaia DR3 1358614983131339392}) has a parallax of $2.69\pm0.01$ mas in \emph{Gaia} DR3, placing it at about 372$\pm$1.4 pc.  Minor corrections to the parallax uncertainty \citep{2021MNRAS.506.2269E} and bias \citep{2021A&A...649A...4L} appear negligible for the $G=9.98$ mag source.  The DR3 parallax places the system about 8.7\% farther than previously estimated by \citet{2017AJ....153..211Z}, which adopted a \emph{Gaia} DR1-informed parallax of $2.92\pm0.23$ mas, including a systematic $-0.325$ mas bias term \citep{2016ApJ...831L...6S}.  The wide companion HAT-P-67B has nearly identical parallax ($2.58\pm0.05$ mas) and proper motions, confirming its interpretation as co-moving, with a projected separation of about 3400 AU.  The IAU naming convention would demand HAT-P-67A\emph{b} to refer to the planet, which orbits the primary.  Hereafter, we simply drop the A designation for notational simplicity since the wide companion will not factor into our analysis.

\subsection{ASAS-SN, DASCH, and ZTF}
We retrieved ground-based photometry with the \emph{All-Sky Automated Survey for Supernovae} (ASAS-SN) using the Sky-Portal \citep{shappee14,2017PASP..129j4502K}.  The precision of ASAS-SN was too low to perceive stellar variability, so we can place a relatively uninformative limit of $<5\%$ stellar variability on years-long timescales.

Similarly, \hatp appears in the Harvard Plate Archive, with 5809 measurements digitized through the Digital Access to a Sky Century at Harvard (DASCH) program, spanning over 120 years of coarse photometric monitoring.  In principle, these datasets could inform long-term variability trends such as stellar cycles.  In practice, the 0.15 magnitude jitter appears too coarse to perceive any genuine astrophysical variability, with no conspicuous trend seen.  We can therefore place a relatively mild constraint that the star appears stable at the $\sim30\%$ level over periods of tens to hundreds of years.

HAT-P-67A was saturated in ZTF \citep{2019PASP..131a8002B}, but the M-dwarf companion HAT-P-67B had up to thousands of visits across several years.  The data quality appeared too poor to perceive any genuine astrophysical variability, with the indication of some lunar background signals in the periodogram.

\section{Analysis} \label{secAnalysis}

\subsection{Gaia DR3}\label{analysisgaiaDR3}
\citet{2017AJ....153..211Z} previously derived stellar radius estimates of 2.1-2.7 $R_\odot$ through Spectral Energy Distribution (SED) and isochrone fitting as part of a joint orbit fit.  We systematically increase those stellar radius estimates by 8.7\% to match the greater Gaia \emph{DR3} distance (\S \ref{gaiadr3}).  For a fixed $R_p/R_\star$ from the measured transit depth, the larger $R_\star$ implies a proportionally larger planet radius. This update systematically decreases the estimate for the already-low density of HAT-P-67 by 28\%, to a mere $<$0.035 g$\;$cm$^{-3}$, albeit with significant uncertainties from the weak mass constraint.  The luminosity increases by 18\%, to about 10.3 $L_\odot$.

\begin{figure}
    \includegraphics[width=\linewidth]{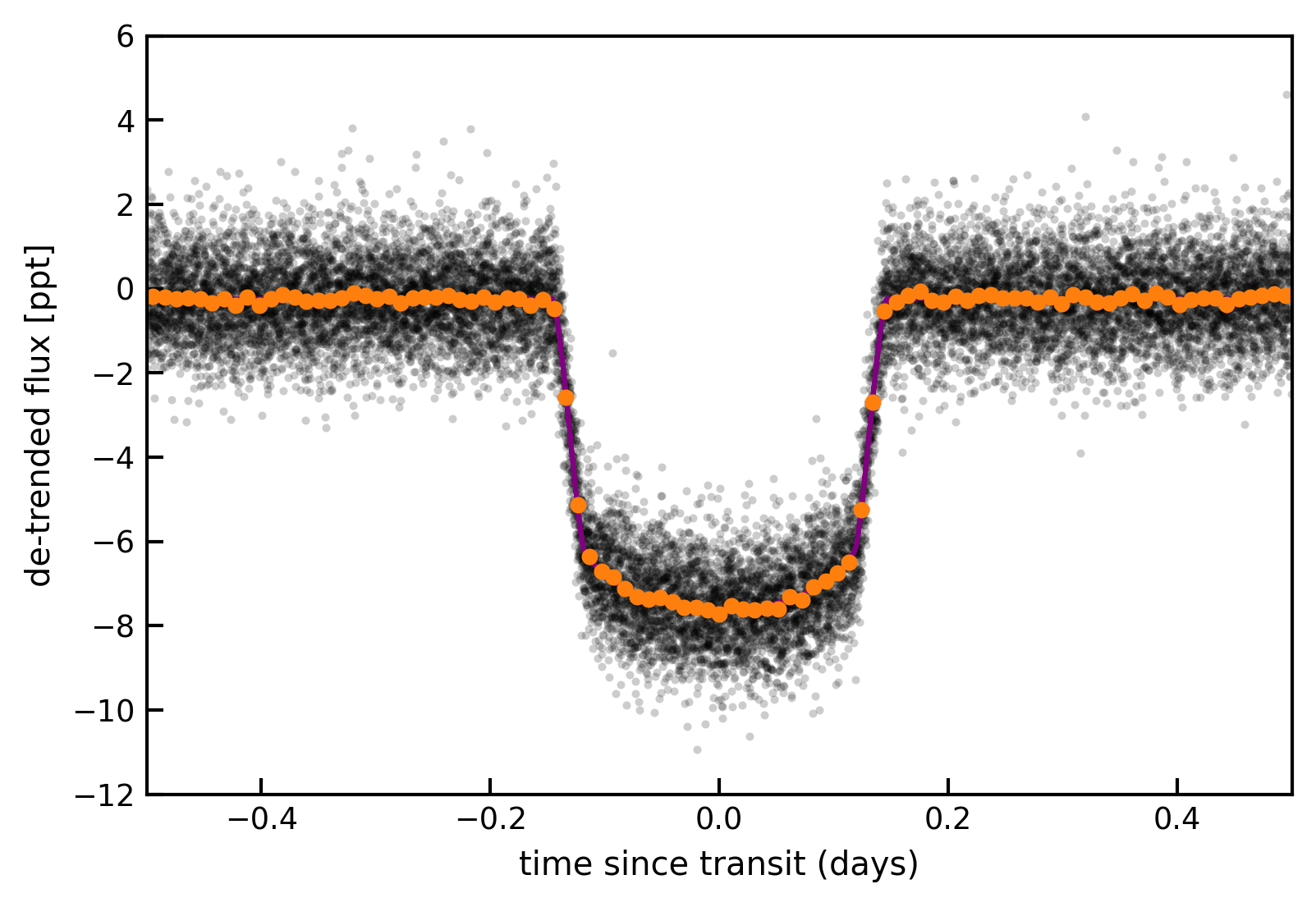}
    \caption{Best fit orbit overlaid on the composite TESS lightcurve.  Revised planet properties are consistent with \citet{2017AJ....153..211Z}, with a systematic shift from a revised \emph{Gaia} DR3 distance.}
    \label{fig:transit}
\end{figure}

\subsection{TESS Light Curve}
Previously, \hatpb~transits had only been detected with \emph{HATNet} \citep{2004PASP..116..266B} and followed up with KeplerCam on the FLWO 1.2 m telescope \citep{2017AJ....153..211Z}. These ground-based photometers were not intended to measure weak, long-term stellar variability signals.  We, therefore, examined the \emph{TESS} lightcurves for out-of-transit photometric variability.  The revised precision and continuous coverage of \emph{TESS} can also refine the orbital solution.

\subsubsection{Revised exoplanet orbital parameters}
We assembled a composite lightcurve by stitching TESS Sectors 24, 26, 51, 52, 53, which were reduced with the default SPOC pipeline \citep{2020RNAAS...4..201C}, and lightly post-processed with \texttt{lightkurve} \citep{geert_barentsen_2019_2565212}.  These TESS data exhibit an RMS scatter of better than 1 part-per-thousand (ppt) at native 2-minute sampling.  We fit a Keplerian orbit model to the TESS lightcurve using the \texttt{exoplanet} framework \citep{exoplanet:joss}.  We obtained revised orbital properties shown in Table \ref{tabOrbit}.  These properties are broadly consistent with the previously reported values from \citet{2017AJ....153..211Z} and the updated ephemeris of \citet{2022ApJS..259...62I}.  Figure \ref{fig:TESSoverview} shows an overview of all the TESS Sectors with a Gaussian Process trendline in green highlighting the out-of-transit modulations, and the exoplanet transit model in orange.  Figure \ref{fig:transit} shows the best-fit orbit overlaid in purple on the detrended TESS lightcurve, which is binned in the orange dots.  Table \ref{tabOrbit} lists only one of the previous orbit determinations---also assuming a circular orbit---compared with the orbital solution reported here.

\subsubsection{Stellar rotation rate from periodogram analysis}\label{TESSmodulation}

The TESS Sector 26 lightcurve exhibits a weak $\sim$3 ppt peak-to-valley out-of-transit modulation. TESS Sectors 51-53 do not show as conspicuous a modulation signal but still show some ostensibly stellar variability.  The ebb and flow of the modulation can be seen in the minimally processed TESS lightcurve in Figure \ref{fig:TESSoverview}.  We fit the TESS lightcurve modulation with a quasiperiodic Gaussian Process (GP) model using \texttt{celerite} \citep{celerite1,celerite2}.  The GP model fit to the entire composite lightcurve yields a 4.7-day period; when fit to individual sectors alone, the periods hover around 5.9 days.  The signal is weak enough, especially in Sectors 51-53, that \emph{TESS} instrumental systematics may contribute some of the out-of-transit modulation that we see in Figure \ref{fig:TESSoverview}.

We can independently constrain the expected stellar rotation period based on the stellar radius estimate, $v\sin{i_\star}$, and the observation of the planet's orbital inclination $i_p\sim90^\circ$ from orbit fitting and Doppler tomography \citep{2017AJ....153..211Z}.  We assume spin-orbit alignment, $i_\star \sim i_p$.  We adopt a high limit of $v\sin{i}=35.8\pm1.1$ km/s and a lower $v\sin{i}=30.9\pm2$ km/s value if macroturbulence is accounted for.  We obtain a range of $3.2 < P_\mathrm{rot}  < 4.8 $ days.  This range is typical for F stars that have not yet evolved too far into the subgiant branch \citep{2022ApJ...930....7A}.  The geometrical constraint comports with the 4.7-day GP-based modulation period derived from the stitched TESS lightcurve but is lower than the 5.9-day per-Sector fits, slightly preferring the lower 4.7-day modulation as the stellar rotation period.  The 5.9-day value would require a stellar radius over 3.2$R_\odot$, which appears implausible.

\begin{deluxetable}{lRR}
    \tablewidth{0pc}
    \tabletypesize{\scriptsize}
    \tablecaption{
        Revised orbital and planetary parameters
        \label{tabOrbit}
    }
    \tablehead{
        \colhead{Parameter}   &
        \colhead{Zhou et al. 2017} &
        \colhead{This Work\tablenotemark{a}}
    }
    \startdata
    ~~~$\rstar$ ($\rsun$)\dotfill      & 2.546_{-0.084}^{+0.099} &  2.65\pm0.12 \\
    ~~~$P$ (days)             \dotfill    & 4.8101025_{-3.3\times 10^{-7}}^{+4.3\times 10^{-7}} & 4.8101046\pm4.7\times10^{-6}  \\
    ~~~$T_c$ (${\rm BTJD}$\tablenotemark{b})  \dotfill    & -1038.61533_{-0.00064}^{+0.00076} & 2694.027\pm0.001 \\
    ~~~$T_{14}$ (hours)  \dotfill    & 6.9888 \pm 0.046 & 7.062\pm0.028 \\
    ~~~$\rpl/\rstar$          \dotfill    & 0.0834\pm0.0017 &  0.08396_{-0.00040}^{0.00035}\\
    ~~~$\arstar$              \dotfill    & 5.691_{-0.124}^{+0.057} & 5.036_{-0.086}^{+0.089}\\
    ~~~$b \equiv a \cos i/\rstar$ \dotfill    & 0.12_{-0.08}^{+0.12} &  0.509_{-0.029}^{+0.026} \\
    ~~~$K$ (\ms)              \dotfill    & <36\,(1\sigma)  & 33_{-15}^{+21}\\
    ~~~RV jit. (\ms) \dotfill    & 59 & 164 \\
    ~~~RV sys.(\kms)\dotfill    & -1.4 \pm 0.5 & -0.07\pm0.03 \\
    ~~~$\mpl$ ($\mjup$)  \dotfill    & 0.34_{-0.19}^{+0.25} & 0.32_{-0.15}^{+0.21}\\
    ~~~$i$ (deg)        \dotfill    & 88.8_{-1.3}^{+1.1} & 84.19_{-0.41}^{+0.43} \\
    ~~~$a$ (AU)         \dotfill    & 0.06505_{-0.00079}^{+0.00273} & 0.062\pm0.003\\
    \enddata
    \tablenotetext{a}{Zero-eccentricity fit}
    \tablenotetext{b}{BTJD $\equiv$ BJD $-$2457000}
\end{deluxetable}

\begin{figure}
    \centering
    \includegraphics[width=0.98\linewidth]{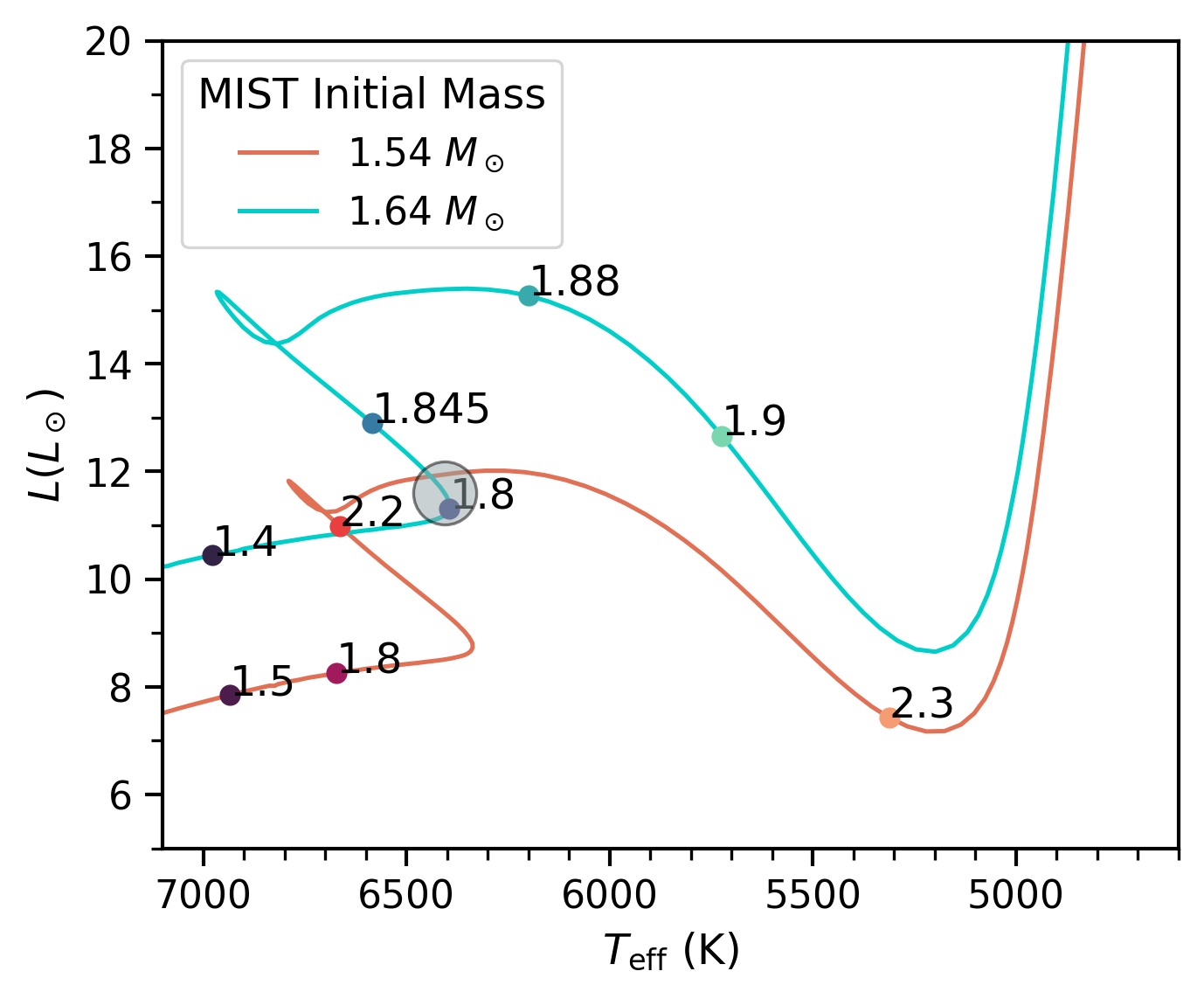}
    \caption{MIST evolutionary model tracks for HAT-P-67.  The $T_\mathrm{eff}$ and luminosity of \hatp (gray circle) are consistent with either the late main sequence or recently evolved subgiant.  The labeled numbers indicate age in Gyr.}
    \label{fig:evolTracks}
\end{figure}

\subsection{Revised system evolutionary state}\label{secMISTtracks}
The slightly increased distance and radius imply an 18\% more luminous host star than previously estimated.  Figure \ref{fig:evolTracks} shows evolutionary tracks using the solar-metallicity MESA \citep{2011ApJS..192....3P,2013ApJS..208....4P,2015ApJS..220...15P} Isochrones and Stellar Tracks \citep[MIST;][]{2016ApJS..222....8D,2016ApJ...823..102C} for HAT-P-67, which has [Fe/H]$=-0.08\pm0.05$ \citep{2017AJ....153..211Z}.  The tracks show that a 1.8 Gyr, 1.64 $M_\odot$ HAT-P-67 could be at the end of its main sequence lifetime (teal track), yielding a gradual rise in incident radiation over the lifetime of \hatpb.  Alternatively, a 1.54 $M_\odot$ evolutionary track would send HAT-P-67 along the subgiant branch at an age over 2.2 Gyr, with a rapid 50\% increase in luminosity, potentially leading to ``re-inflation'' of the planet \citep{2021ApJ...909L..16T}.  Either evolutionary track appears consistent with the observed SED, stellar surface gravity, and available constraints from spectral fitting.  We, therefore, adopt a MIST-based 2.0$\pm$0.2 Gyr age for the system, in-between the previous Geneva and Dartmouth-based scenarios \citep{2017AJ....153..211Z}.  Additional metallicity and rotation effects could slightly alter the evolutionary tracks and therefore increase uncertainty in inferred ages and masses.

\begin{figure}
    \includegraphics[width=\linewidth]{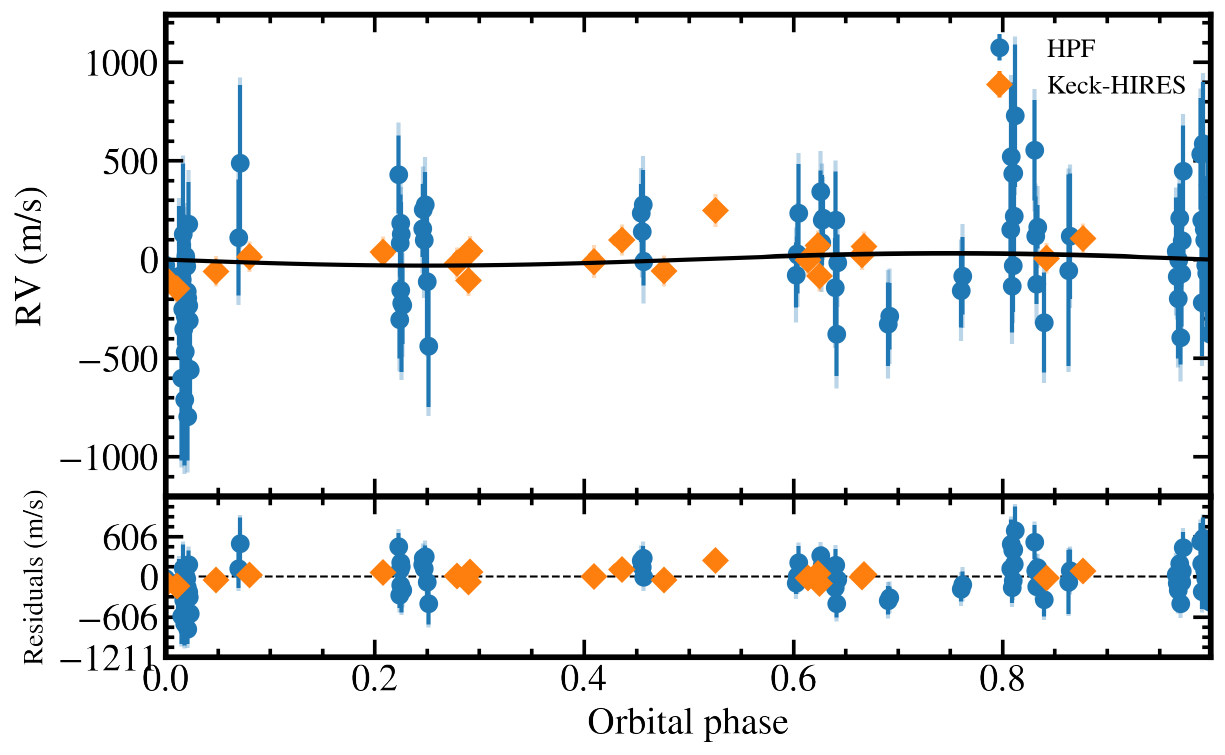}
    \caption{RV orbit fit including both HPF and Keck HIRES data points.  The RV information content in HPF is much less than in Keck HIRES for this F5 spectral type, and so the joint fit constrains the mass to roughly a Saturn mass.}
    \label{fig:RVfit}
\end{figure}

\subsection{HPF analysis I: RV fitting}\label{secRVfit}

We conducted orbit fitting via precision radial velocity (PRV) measurements following the procedures described in \citet{2021AJ....161..173T} and updated for joint lightcurve and RV fits \citep{2022AJ....163..225T}, but without a joint Gaussian Process modeling procedure \citep{2023ApJ...950..162T}.  Here we included the new TESS lightcurves and the existing Keck HIRES points reported in Table 3 of \citet{2017AJ....153..211Z}.  The HPF points exhibited an RV jitter of $\sim164\;\ms$, a few times larger than the typical Keck HIRES measurements, due to the lower information content in the near-IR than the visible for this relatively rapidly rotating F star.  So even though the HPF points were more numerous, their marginal value for RV orbit determination was subdued.  The period and $T_0$ were fixed from the \emph{TESS} transits, and we assume a circular orbit with $K>10$ \ms.  The semiamplitude prior acts as a mass constraint, excluding planets with masses so low that the observed radius would exceed the Hill radius (Roche lobe overflow).  Figure \ref{fig:RVfit} shows the fitted RV semiamplitude $K=33_{-15}^{+21}$ \ms, a relatively weak constraint but consistent with broadly Saturn-mass planets.  Table \ref{tabOrbit} lists the revised properties, with minor differences from the existing values.  The semiamplitude prior accounts for the apparent improvement in the mass constraint.  Hypothetically, a truly Roche lobe overflow planet could be consistent with the available data, so our mass constraint could be considered an upper limit.

\subsection{HPF analysis II: \ion{He}{1} 10833 \AA} \label{secHeAnalysis}
The preparation for analyzing the Helium feature followed a similar but slightly different procedure than that used for the RV. Namely, we used \texttt{Goldilocks}\footnote{\url{https://github.com/grzeimann/Goldilocks_Documentation}} for 2D \'echellogram reduction.  This tool outputs 1D extracted spectra for the target and two reference fibers: blank sky and a laser frequency comb \citep[LFC,][]{2019Optic...6..233M}.  The observations were acquired with the LFC turned off, so this unilluminated spectrum was discarded.

The sky fiber and target fiber have slightly different throughputs and illumination properties, with the target fiber receiving $93\%$ of the flux of the sky reference fiber on average.  This ratio depends on wavelength and season at the few percent level.  We quantified the wavelength dependence by acquiring calibration observations of blank sky in both the target and sky fiber.  We applied this wavelength-based scale factor to each target spectrum's associated reference sky fiber to achieve sky-line subtraction residuals typically less than the photon noise \citep{2022JOSS....7.4302G}.  A few lines exhibit residuals that may arise from genuine differences in the local atmospheric conditions between the target and sky fiber.

\begin{figure}
    \includegraphics[width=\linewidth]{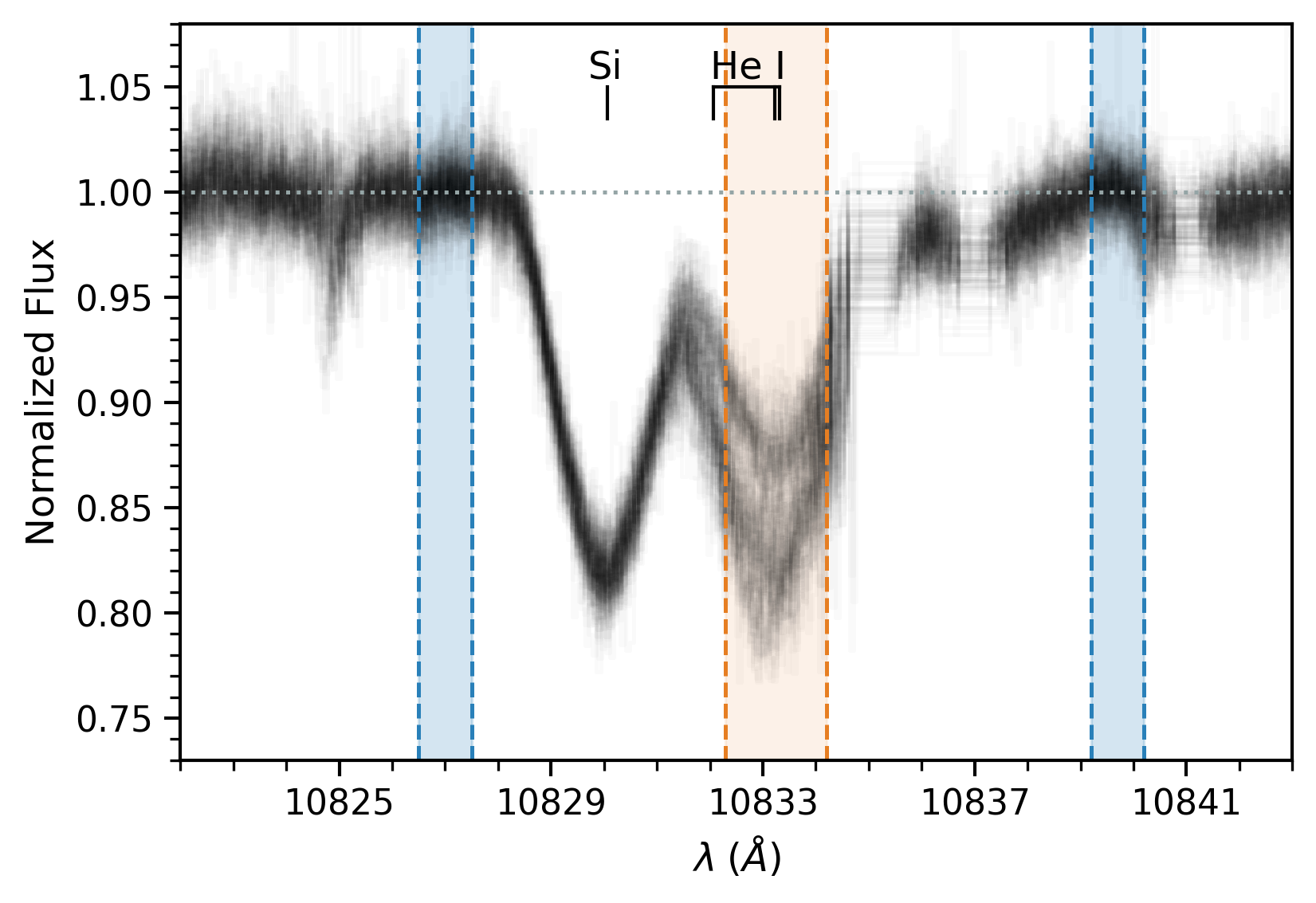}
    \caption{Overlay of all 152 individual HPF exposures of HAT-P-67, spanning 2020-2022. Variability is seen in the \ion{He}{1} 10833 \AA~ triplet near the vertical orange shaded band but not in the adjacent Si line.  These snapshot spectra were barycentric corrected and continuum flattened with a linear fit to the regions in the blue vertical bands.  Sharp telluric absorption lines have been masked in regions near 10835 and 10837.5 \AA.}
    \label{fig:HPFheliumOverview}
\end{figure}

The HET's fixed-altitude design means that the airmass remains relatively constant across all pointings, lowering the telluric line variability compared to fully steerable telescopes, which may sample a wider range of airmasses.  Variability in atmospheric conditions still makes it difficult to mitigate telluric lines to within the photon noise limit, so we masked spectral regions predicted to have significant telluric lines with a template generated by \texttt{telfit} \citep{2014AJ....148...53G}.  We then shifted the spectral coordinates to their common barycentric-corrected reference frame \citep{2014PASP..126..838W}, as implemented in \texttt{astropy} \citep{2013A&A...558A..33A,2018AJ....156..123A,2022ApJ...935..167A}.  The spectral continuum was flattened and normalized to two pre-selected continuum indices highlighted as vertical blue bands in Figure \ref{fig:HPFheliumOverview}.  The sky-subtraction, telluric masking, barycentric correction, flattening, and all of our other standard pre-processing steps were carried out in the open-source Python interface \texttt{muler} \citep{2022JOSS....7.4302G}.

Figure \ref{fig:HPFheliumOverview} shows a zoom-in on the He 10833 \AA~ region of interest, with all 152 individual exposures overlaid.  We see variability in the Helium line of up to 10\%, much greater than the $<1\%$ pixel-to-pixel variation.  The feature width spans about 3~\AA.

\begin{figure}
    \includegraphics[width=\linewidth]{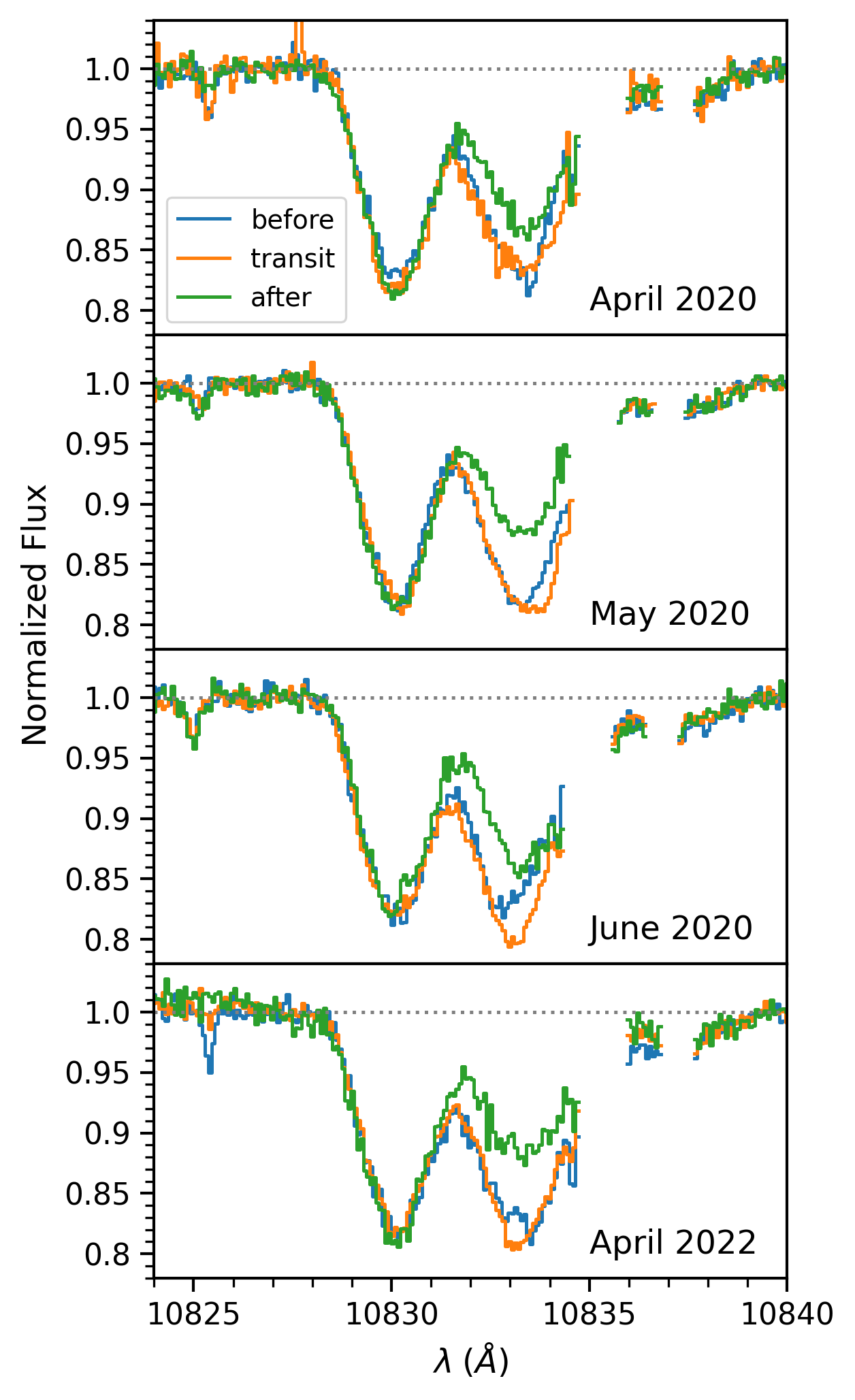}
    \caption{Four observing campaigns centered on an in-transit epoch with before-and-after visits typically separated by 1 night. The after-transit spectra tend to show negligible absorption.  Line-of-sight velocity substructure can be seen in the before and during transit epochs.}
    \label{fig:HPFperCampaign}
\end{figure}

Figure \ref{fig:HPFperCampaign} shows the spectra for four campaigns, with before and after spectra showing conspicuous excess absorption during transit and 1 day before transit but negligible excess absorption after transit.  The individual transits show significant morphological variation, with substructure in the bulk line-of-sight velocity distribution.

\begin{figure}
    \includegraphics[width=\linewidth]{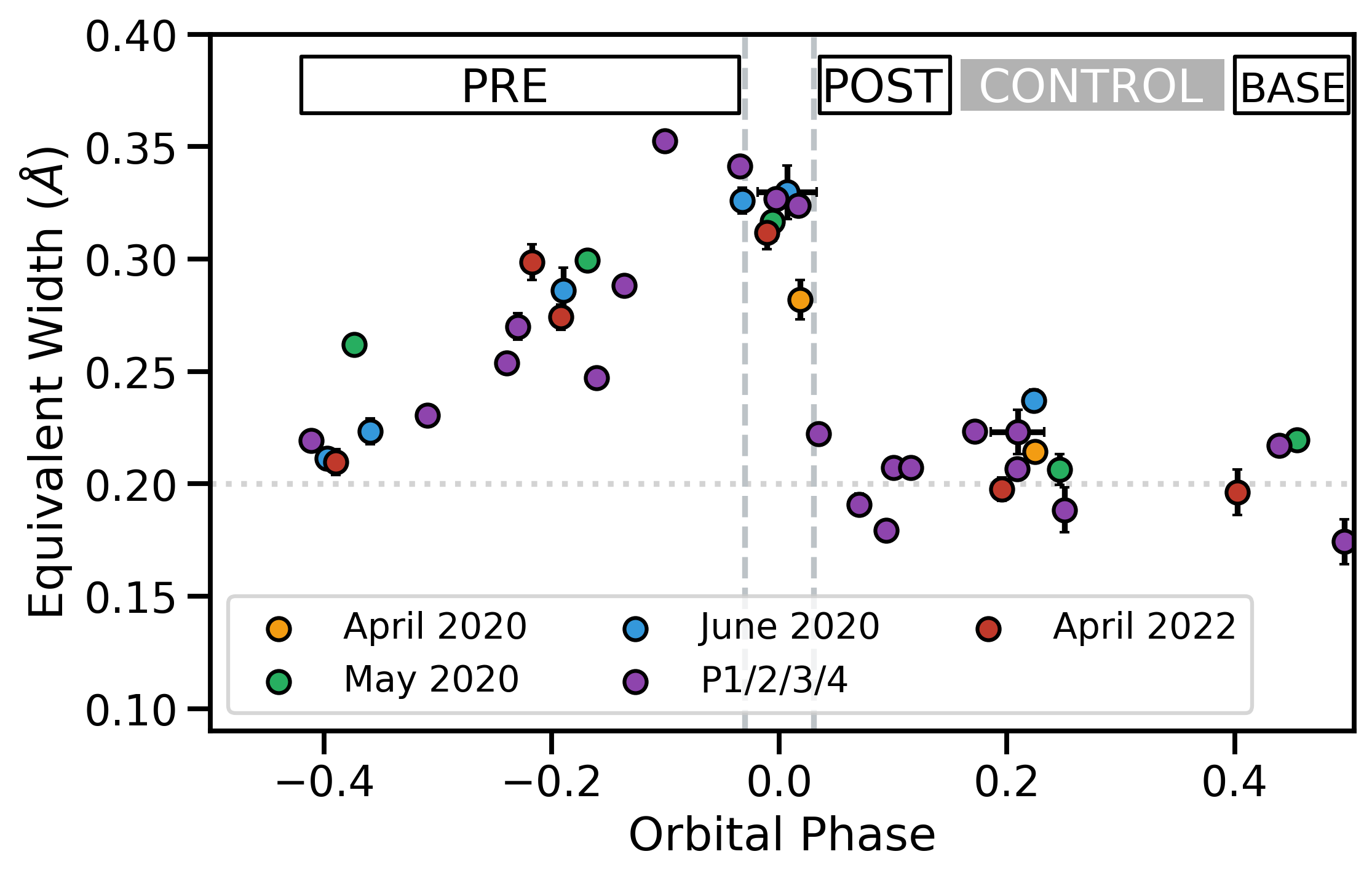}
    \caption{Equivalent Width lightcurve of Helium absorption in HAT-P-67b.  The system exhibits a characteristic absorption leading up to transit, followed by a sharp decline after transit passage.  The EWs were computed in the orange shaded band (10832.3$-$10834.2 \AA) in Figure \ref{fig:HPFheliumOverview}.}
    \label{fig:HPFtimeseries}
\end{figure}

\begin{figure}
    \includegraphics[width=\linewidth]{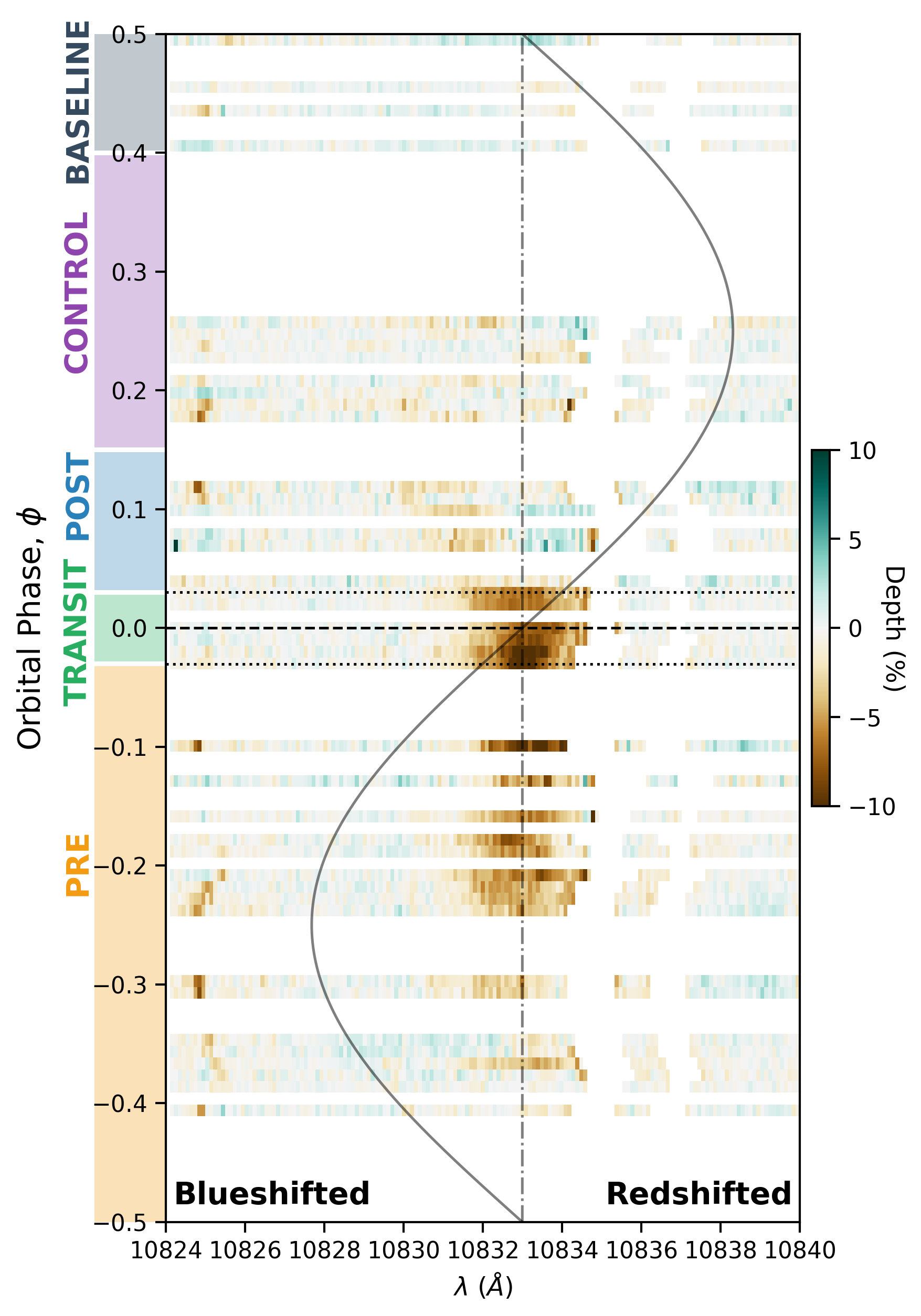}
    \caption{Absorption depth phase scan over the entire HAT-P-67b orbit from 41 visits with HPF.  The planet orbital rest frame is shown as the black sine wave. Gaps in data at 10834-10836 \AA~ arise from telluric masking.  Unmasked telluric lines are perceptible at 10825 \AA.}
    \label{fig:HPFscanResid}
\end{figure}

The 13.8 hours of exposure, combined with \object{HAT-P-67 b}'s short ($P\sim4.8$ day) period, means that a large fraction of the orbit has been collected, with some phases (especially near transit) heavily sampled, and some other out-of-transit phases sampled more sparsely. The largest gap spans merely 0.18 in phase, as seen in the Equivalent Width time series in Figure \ref{fig:HPFtimeseries}.  For visualization purposes, we constructed a 2-D intensity phase scan, binning in phase and wavelength, and spanning the entire orbit of \object{HAT-P-67 b}.  Figure \ref{fig:HPFscanResid} overlays the planet's approximately $\pm150\;$km/s orbital Doppler velocity, with the stellar rest frame velocity demarcated by the vertical line.  The vanishing $<36$ m/s reflex motion of the star is imperceptible at this scale.  The horizontal dashed lines indicate the moments of transit ingress (-0.03), mid-center (0.0), and egress (+0.03), demarcated as \texttt{TRANSIT} in the figure.

We define 4 additional distinct groups of phases with adjectival qualifiers in anticipation of the need to discuss bulk trends: \texttt{PRE}, \texttt{POST}, \texttt{CONTROL}, and \texttt{BASELINE}.  The \texttt{BASELINE} phases define the out-of-transit baseline.  The ``control'' spectra designate phases just before the baseline but are not used to define the baseline, to inspect minor correlation structure in the spectra without dividing it out.  Table \ref{tabHPFLog} lists the adjectival qualifiers for each HPF visit.

The in-transit phases exhibit over 10\% excess absorption peaks.  A large absorption signal can be seen preceding transit ingress, with some significant absorption evident before $-0.2$ in phase.   The egress drop-off is extremely sharp, with almost no excess absorption directly after transit, as seen in the Equivalent Width (EW) time series (Figure \ref{fig:HPFtimeseries}).  We construct a fractional residual absorption spectrum by subtracting off and re-normalizing to the non-varying baseline.  We define the non-varying baseline spectrum as the average over phases $0.4-0.5$, which exhibited stable spectra with the least absorption.  Figure \ref{fig:HPFscanResid} shows that significant absorption can be seen as a few percent at $-0.37$ in phase, rising to 10\% just before and during transit. The abrupt dropoff in helium excess at planet egress is conspicuous.

The underlying structure of the metastable \ion{He}{1} triplet consists of three quantum components, two of which ($J=$ 1 and 2) are typically blended owing to Doppler broadening from the finite temperature of the gas $T_0$.  HAT-P-67~b exhibits blending of all three components ($J=$ 0, 1, and 2) into a single broad Gaussian-like feature seen in Figure \ref{fig:HPFheliumOverview}.  The mere observation of this broadening implies that HAT-P-67~b probes a larger velocity dispersion than typical measurements and may be associated with either a higher gas temperature, complex planetary wind flows, or some mix of both.  Whatever the cause, we treat the feature as a single Gaussian line for the purpose of estimating bulk characteristics of the Helium excess absorption feature.  We restricted the fitting to the 10828$-$10838 \AA~region of the residual spectra.  The model constructed in this way has a total of 4 parameters: amplitude $A$, location $\lambda_c$, Gaussian width $\sigma$, and constant offset.  We repeated a similar process for the nearby Si line at 10830 \AA~as a control sample.

Excess is confidently detected from phases $-$0.3 to $+$0.03.  The full-width at half maximum (FWHM) of the feature is about 1.8$-$2.6~\AA, equivalent to a line-of-sight velocity distribution in the range of 50-75~km/s.  Individual fits to the He feature show a typical line center uncertainty $\pm0.1$~\AA~ or better.  The out-of-transit phases exhibit a line center position of 10833.4~\AA, consistent with the stellar \ion{He}{1} rest frame zero velocity.  The excess absorption resides slightly blueward of the stellar zero restframe velocity reference.  As mentioned previously, the telluric masking near 10836 \AA~ censors our view of the existence (or not) of any redshifted lobe in this region.  The excess detections exhibit a bulk blueshift of up to 30~km/s.

The excess absorption spectrum exhibits an increasing line-of-sight velocity distribution as the planet transits, going from 60~km/s at ingress, up to 100 km/s at egress.

Finally, we detect a much weaker but still significant post-egress blue-shifted absorption.  This lobe appears to decrease linearly in wavelength, corresponding to a characteristic blueshift increasing from near zero at transit midpoint to 80~km/s at 12 hours after transit.  It exhibits a slightly lower line-of-sight velocity distribution of 50 km/s, about half of the peak during transit.

We examined other spectral lines using the same methodology as above, finding no detectable variability in the \ion{Ca}{2} infrared triplet, Hydrogen lines, or other neutral metal lines.  We show some of the diagnostic plots for these non-detections in Appendix Section \ref{appendixSec}.

\subsection{Keck HIRES}
We retrieved 22 epochs of archival \emph{Keck HIRES} spectra via the Keck Observatory Archive (KOA). Of these, 19 were acquired through the iodine cell as reported by \citet{2017AJ....153..211Z}, providing RV orbit constraints.  The 14\arcsec~ slit height of the C2 decker appears to cause order overlap in \ion{Ca}{2} H and K lines, which contaminated the spectral extraction for 14 of the 22 spectra.  The 3\farcs5 B2 decker allowed the faithful extraction of this region for 6 spectra, which exhibit no perceptible \ion{Ca}{2} H and K variability.  The H$\alpha$ region exhibited no perceptible variability, but the timing of these spectra coincidentally missed the orbital phases ($-0.3$ to $+0.03$) where we see the greatest \ion{He}{1} 10833 \AA~ absorption excess, leaving open the possibility that detectable H$\alpha$ absorption could be present at more favorable orbital phases.  New spectra or a more careful extraction of the archival C2 decker spectra would be needed.

\subsection{Velocity substructure}
\begin{figure}
    \centering
    \includegraphics[width=\linewidth]{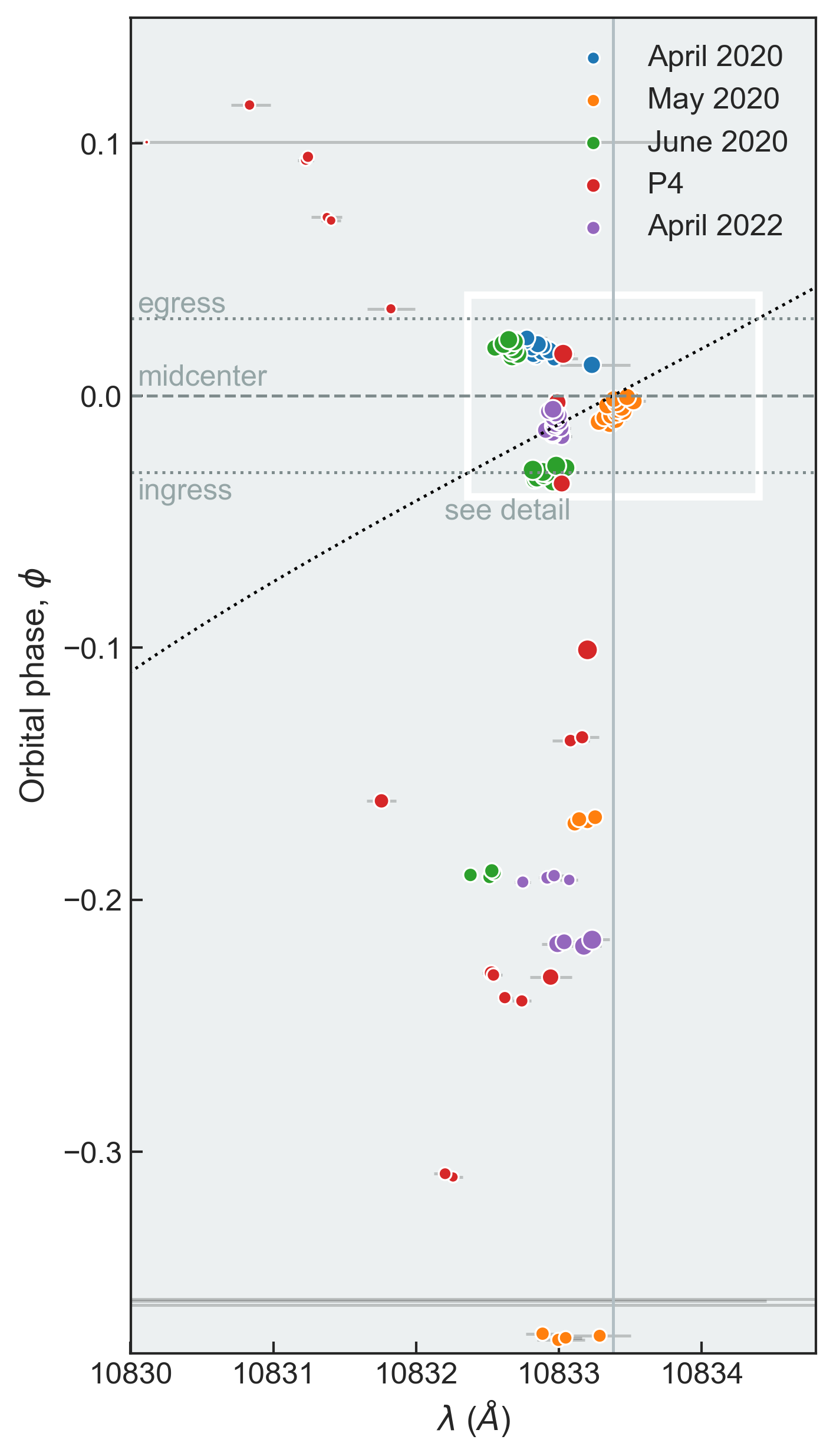}
    \caption{Line centroid positions of the  \ion{He}{1} 10833 \AA~ feature.  A slight blueshift relative to the stellar restframe (vertical line) can be seen, with a weak excess blueshift trend after planet egress. The line-of-sight velocity distribution of 2-3 \AA~ means that the gas distribution exhibits both advancing and receding lobes at all phases other than post-egress.}
    \label{fig:centroids}
\end{figure}

Figure \ref{fig:centroids} shows the centroid positions of the Helium excess feature.  The bulk velocity drifts of the planet's outflow can be seen with a few conspicuous trends.  The pre-transit phases ($-0.3$ to $+0.03$) show a slight tendency towards blueshift relative to stellar restframe, with an accelerating blueshift from phases $-0.3$ to $-0.15$, albeit with some visit-to-visit scatter.  The post-transit observations ($+0.03$ to $+0.15$) exhibit weak-albeit-significant equivalent widths, as denoted by their smaller marker size.  The centroids dramatically accelerate to larger bulk blueshifts, as low as 10831 \AA.

At the moment of transit midcenter, the centroids exhibit a redshifted absorption relative to the planet restframe. Measurements from April 2022 and May 2020 transits partially overlap in phase coverage while yielding slightly different line centroid locations.  Both time series sequences appear consistent with near-zero bulk restframe velocity but with significant variability in the line profile substructure.  The May 2020 sequence shows a slight trend towards alignment with the planet restframe.  The April 2020 and June 2020 partial transits sampled phases close to planet egress, with a clear blueshift relative to both the star and planet.

Overall this pattern of velocity centroids appears consistent with the majority of the gas rapidly obeying the stellar restframe but with a large spread in the line-of-sight velocity distribution, as we would expect from a range of launch angles.  We revisit the interpretations and caveats for this substructure in the next sections.

\section{Atmospheric Escape}\label{secResults}

\subsection{Signal Inconsistent with Stellar Activity Interpretation}
The planet's orbital period is close to the stellar rotational period, so the prospect of stellar activity contamination arises.  Overall, the HPF spectra disfavor a stellar activity interpretation for a few reasons.  First, a hypothetical stellar origin of Helium 10833 \AA~ variability should be accompanied by variability in other tracers.  Instead, the \ion{Ca}{2} Infrared Triplet (IRT) shows stable line profiles in our HPF spectra.  The Appendix Section \ref{appendixSec} discusses these and other non-detections.  Second, the velocity substructure and phasing appears inconsistent with stellar activity: the gradual rise and then abrupt truncation of the Helium excess at the moment of planet egress (Figure \ref{fig:HPFtimeseries}) appears inconsistent with a more smoothly varying stellar variability.  Finally, the weak post-egress excess appears blueshifted to wavelengths as short as 10830 \AA, which would fall entirely outside of the original Helium spectral line's rotationally broadened profile, ruling out a heritage from the star's surface.  

The recent discovery of giant tidal tails in HAT-P-32~b \citep{doi:10.1126/sciadv.adf8736} provides an additional anchor.  HAT-P-32~b is in many ways the most analogous known system to HAT-P-67~b, having comparably low mass and a comparably large radius for similar insolation around an F star.  In HAT-P-32~b the planetary interpretation is unambiguous owing to the different stellar rotation and orbital periods.  In Section \ref{secLackofSaturns} we show that radius inflation mechanisms expect similar and large mass loss rates for both of these systems.

\subsection{Size and geometry of outflow}
Helium absorption occurs predominantly before the planet is in-transit, indicating a \emph{leading} tail escaping the planet.  The leading tail is detected 0.1 AU away from the planet, or 130 planetary radii, well outside of the Roche lobe at $<2.6$ planetary radii.  We are detecting Roche lobe overflow, escaping material that is entirely unbound from the planet.  The bulk of the detectable unbound gas may be understood as following its own Keplerian trajectory at a slightly shorter period orbit, pulling away from the planet parallel to the orbital path.  Such a scenario may arise from preferential emission on the planet dayside.

The large size of the gas stream overfills the transit chord, which means both the advancing blue side and receding red side of the rotating star are nearly continuously obscured in this spin-orbit-aligned system \citep{2017AJ....153..211Z}. If the planet transits near the stellar equator \citep{2017AJ....153..211Z}, the transit chord covers about 10$\%$ of the projected stellar surface area.  Hypothetically a nearly 100$\%$ opaque optically thick stream of gas covering only the transit chord could reproduce the observed 10$\%$ absorption excess.  Alternatively---and more likely---an optically thin gas stream has some additional vertical extent perpendicular to the plane of orbit, which would act to overfill the entire projected stellar disk.  Assuming the gas has a spatially uniform optical depth of about 0.1 would reproduce the 10\% absorption excess.

The leading tail geometry seen in HAT-P-67~b contrasts with the morphology of the HAT-P-32~b system, which exhibits giant symmetric tidal tails with comparable absorption depth preceding and trailing the planet \citep{doi:10.1126/sciadv.adf8736}.  These systems represent two of the most extended features associated with exoplanets, and we compare and contrast them in Section \ref{secDiscuss}.

\subsection{1D Parker wind models and mass loss rates}\label{pwinds}
We explore one-dimensional (1D) models, which offer ease of interpretation and rapid calculation.  We employ the open-source \texttt{p-winds} code \citep{2022A&A...659A..62D}, which implements a transonic 1D Parker wind model with radiative transfer of the Helium 10833 \AA~ triplet \citep{2018ApJ...855L..11O,2020A&A...636A..13L}.

\begin{figure}
    \includegraphics[width=\linewidth]{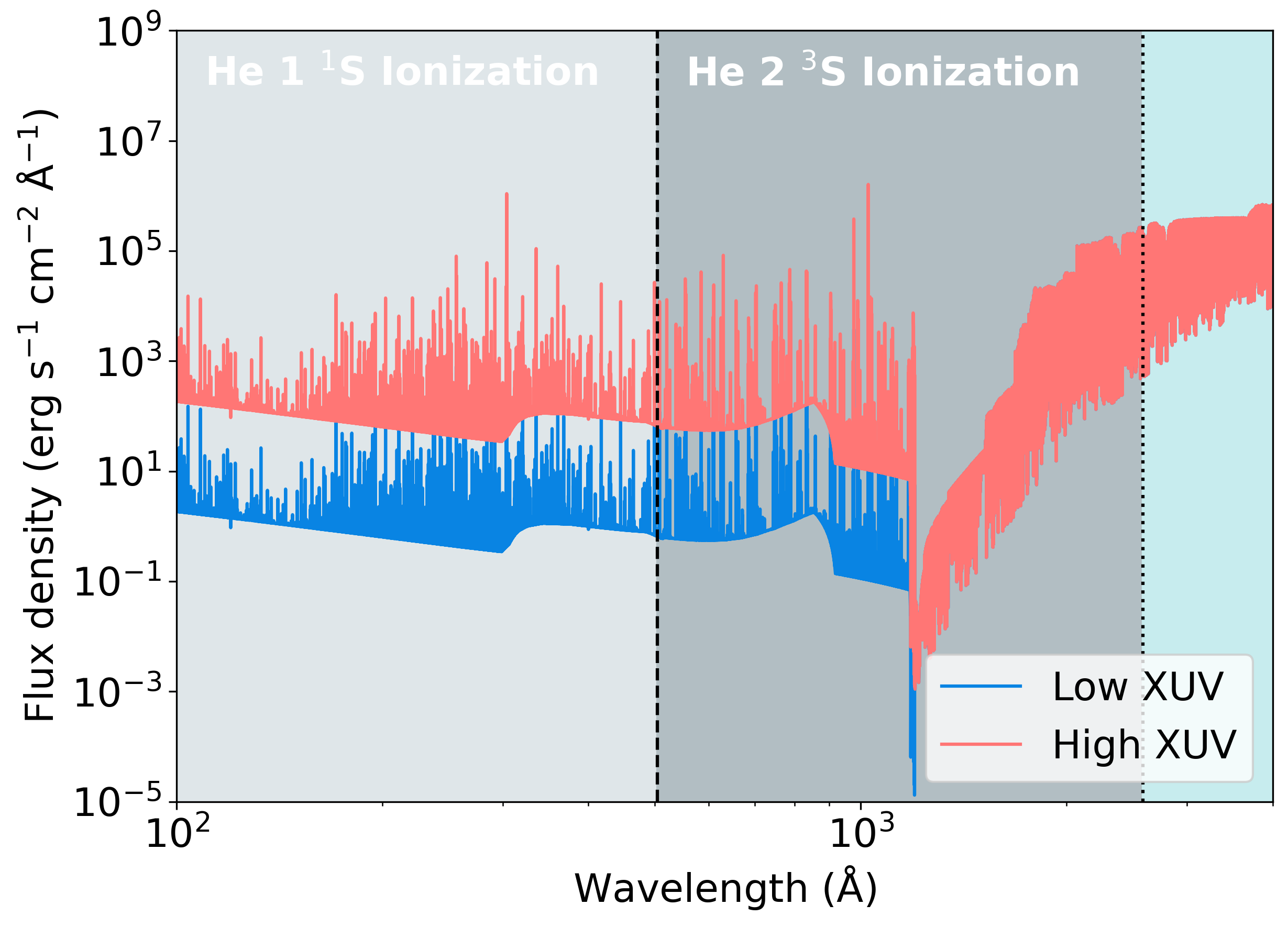}
    \caption{Synthetic XUV radiation scenarios for \hatpb.  The low and high XUV scenarios correspond to $L_X/L_\mathrm{bol}$ of $10^{-6}$ and $10^{-4}$, where $L_X$ is defined as He 1 $^1$S Ionizing photons that trigger the recombination cascade needed for observing the He 2 $^3$S ionization metastable state.  He 2 $^3$S Ionizing photons depopulate the state and suppress the observability of \ion{He}{1} 10833 \AA.  The XUV flux of HAT-P-67 is uncertain.}
    \label{fig:XUV}
\end{figure}

The predicted Helium ionization depends sensitively on the XUV flux \citep{2019ApJ...881..133O}.  Ideally, we would have a measurement of the XUV spectrum of HAT-P-67, but the limited facilities and distance of the source preclude these challenging observations.  Instead, we constructed panchromatic X-ray to visible SEDs by scaling and stitching together synthetic spectra to provide a range of high and low $L_X/L_\mathrm{bol}$.  We chose a 6400 K PHOENIX \citep{husser13} solar metallicity photospheric spectral template with $\log{g}=3.5$---the closest PHOENIX grid point to published and updated values available--- scaled to the solid angle seen by \hatpb.  The F7IV-V \object[tau Boo]{$\tau$ Boo} serves as the closest analog with available synthetic X-ray coronal spectra \citep{2011A&A...532A...6S}.  But \hatp resides closer to the Kraft break than does \object[tau Boo]{$\tau$ Boo}, and the move to hotter F stars may be associated with weakening of the stellar wind and other atmospheric changes leading to the prospect of much lower XUV luminosity for HAT-P-67 \citep{2022ApJ...930....7A}.  So there remains significant uncertainty about the applicability of the \object[tau Boo]{$\tau$ Boo} XUV spectrum to HAT-P-67.  We scaled the synthetic SED of \object[tau Boo]{$\tau$ Boo} retrieved from X-exoplanets \citep{2011A&A...532A...6S} so that the integral of ground-state ionizing photons ($\lambda<504\;$\AA) exhibited $L_X/L_\mathrm{bol} \in [10^{-6}, 10^{-4}]$.  Figure \ref{fig:XUV} shows two conceivable SEDs with these high and low radiation hardnesses.

The SED shows that the XUV corona model does not extend redward to 2600 \AA~photons capable of ionizing out of the 2 $^3$S Helium metastable state.  This apparent deficit should be negligible since the F spectral type obtains most of its NUV photons ($\lambda\sim2600$ \AA) from the Wein side of the photospheric spectrum, so the hardness of the radiation stems mostly from the assumptions of the corona flux level.  The wavelength-dependent cross section for absorption (not shown) is largest just blueward of the ionization thresholds, so the differences in the spectrum between 504 and 1000 \AA~ have relatively little impact on the overall ionization out of the metastable state.

\begin{figure}
    \includegraphics[width=\linewidth]{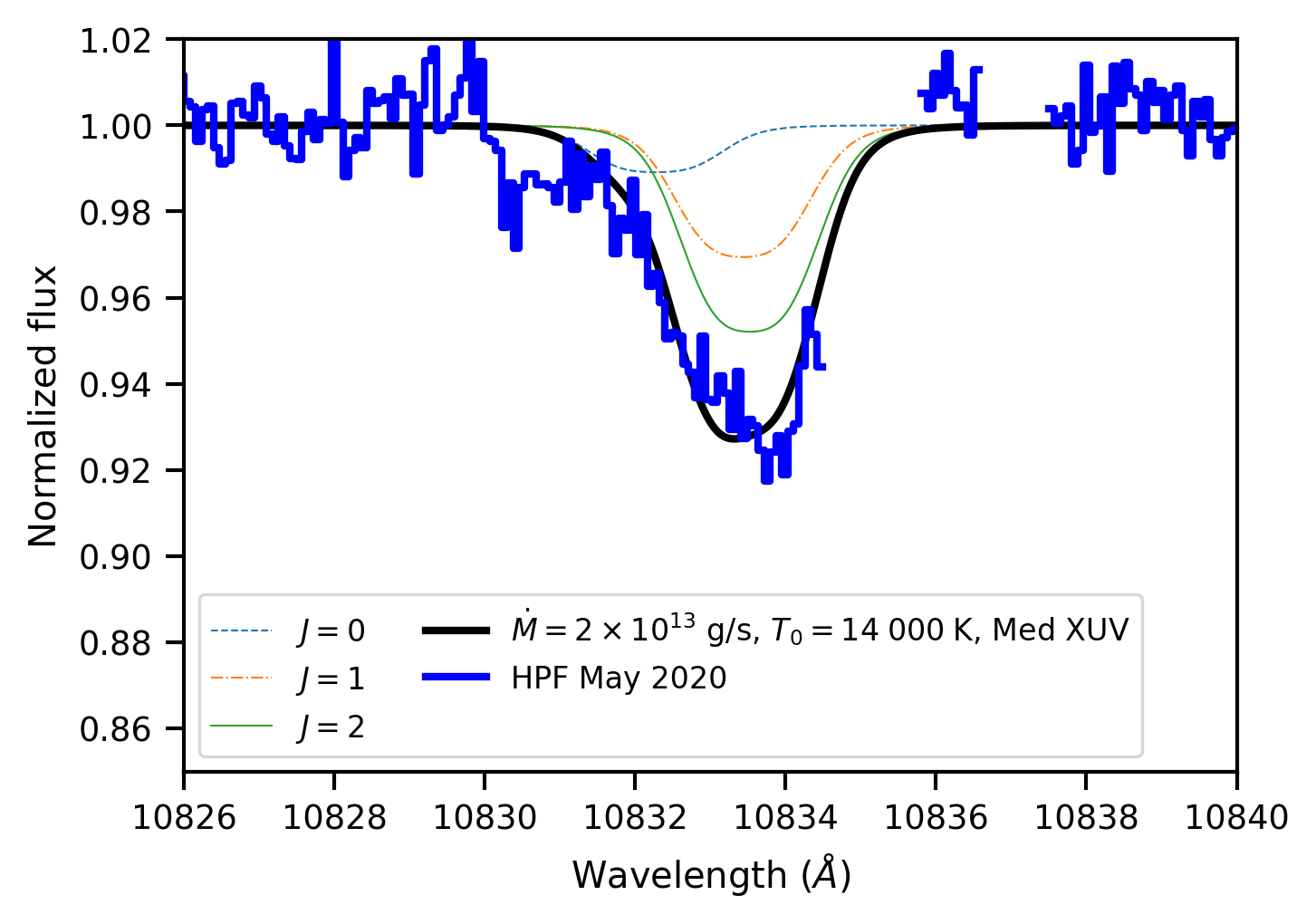}
    \caption{Simulated 1D Parker wind model of Helium absorption in \hatpb.  The spectrum was generated with a mass loss rate of $2\times10^{13}$ g/s, exosphere gas temperature of $14\,000\;$K, and with an XUV luminosity $L_x/L_\mathrm{bol}=10^{-5}$ intermediate between those shown in Figure \ref{fig:XUV}.  The \texttt{p-winds} model yields the \ion{He}{1} 10833 triplet lines (colored thin lines), with a cumulative feature in black resembling the data, shown as the May 2020 in-transit mean.  Additional 3D velocity dispersion of the gas can explain differences between the line shapes of the model and data.  Overall the 1D models may be too simplistic to represent the inherently 3D structure of the outflow.}
    \label{fig:pwinds}
\end{figure}

Equipped with these two SEDs of differing radiation hardness, we explored different mass loss rates and exosphere temperatures.  Figure \ref{fig:pwinds} shows one example model spectrum with $\dot{M} = 2\times 10^{13}$ g/s, and $T_0=14\,000$~K, with an XUV spectrum possessing $L_x/L_\mathrm{bol}=10^{-5}$.  This mass loss rate would imply a characteristic lifetime of $<$1000 Myr.  The model spectrum exhibits approximately the same equivalent width as our HPF observations, making it a hypothetical scenario among a family of partially degenerate solutions.

At least a few shortcomings limit the applicability of this 1D Parker wind model.  First, the model-dependent line profile (\emph{i.e.} width and depth) is degenerate with XUV flux, $\dot{M}$, and $T_0$, and so a range of these parameters can be fine-tuned to obtain a large range of mass loss rates consistent with the data and our limited understanding of the XUV flux.  These known degeneracies have been pointed out previously \citep{2022AJ....164..234V,2019ApJ...881..133O}, but the problem for \hatpb~ appears somewhat more acute since XUV data are scarce for F-type stars near the Kraft break.  Second, the 1D model breaks down when attempting to explain the inherently 3D leading tail geometry.

\subsection{Direct Evidence for Preferential Dayside Mass Loss} \label{secLeading}
Several physical phenomena could conceivably control the geometry and extent of the escaping material.  The stellar potential controls the overall geometry through tides and the Coriolis force, resulting in lobe morphologies that lead and trail the planet \citep{2019ApJ...873...89M}.  Here we explore the three predominant dynamical effects: orbital shear, stellar wind confinement, and day-/night- side mass loss asymmetries.

\begin{figure}
    \includegraphics[width=0.8\linewidth]{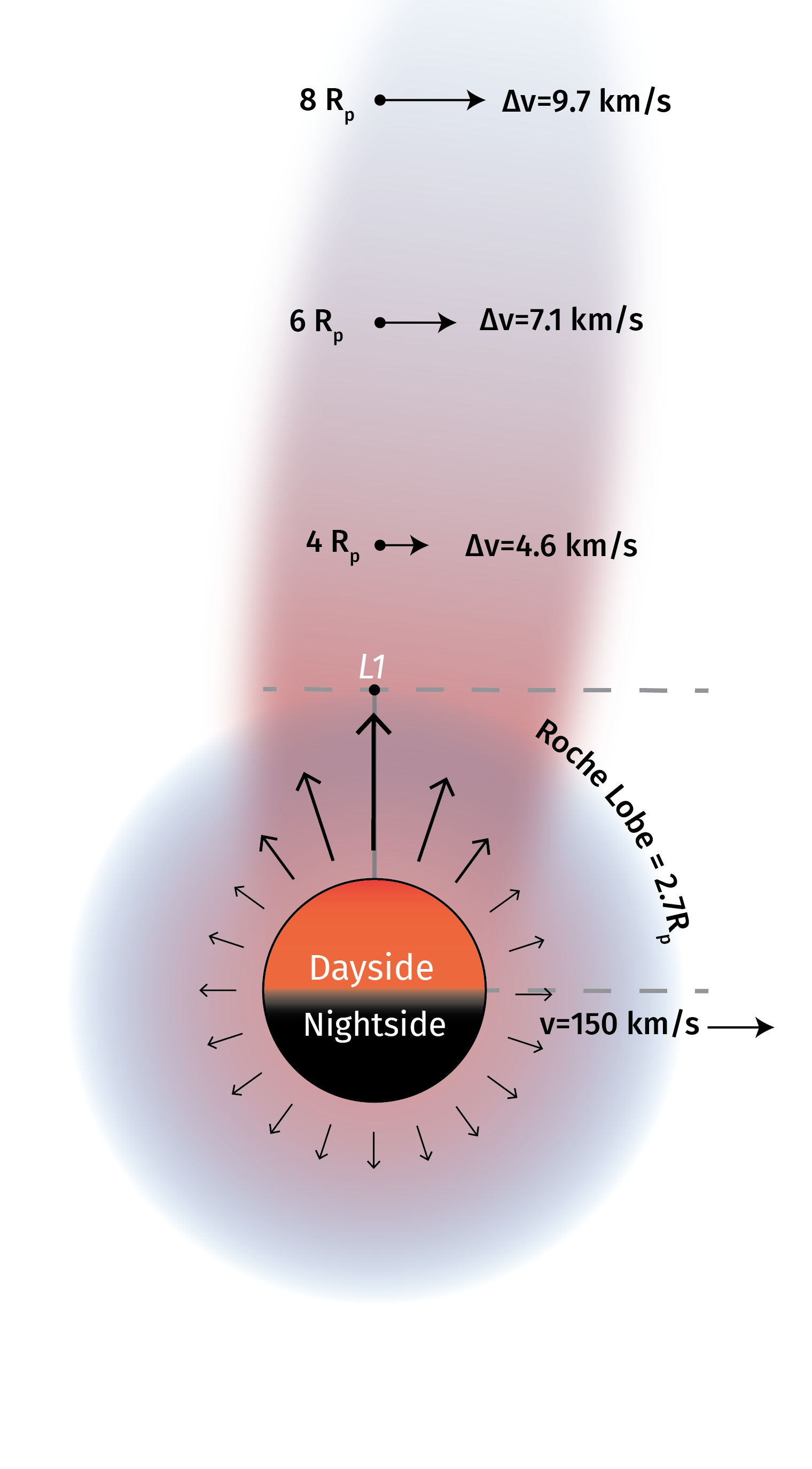}
    \caption{Schematic of Keplerian orbital shear.  Locations outside the Roche lobe experience orbital shear from the Keplerian potential.  A wind launched primarily from the dayside will tend to form a leading tail.}
    \label{fig:KeplerianShear}
\end{figure}

Figure \ref{fig:KeplerianShear} shows an illustration of Keplerian shear adapted to the system properties of \hatpb.  In this shear-dominated scenario, the planetary wind launches primarily from the dayside, with relatively little or no wind launched from the nightside.  The planet wind initially launches radially outward from the exobase, with the strongest wind located near the sub-stellar point, the line connecting the planet to the star along the vertical axis in the figure.  Inefficient grazing incidence heating near the terminator subdues the mass loss in the $x-$direction, meaning that this wind exhibits not a hemispherical shape, but more concentration along the star-planet line.  The gas increasingly experiences the star's Keplerian potential past the Roche lobe, accelerating in the direction of orbital motion, $+x$.  The accelerating column eventually overtakes the planet completely, with the prospect of extending to hundreds of planetary radii.  The mass loss has to be large enough to shield the material to make it observable in the metastable \ion{He}{1} 10833 \AA~triplet.

The observation of such a prominent leading tail leads us to the inescapable conclusion that HAT-P-67~b is predominantly losing mass on the planet's dayside.  An isotropic mass loss would manifest a comparably conspicuous trailing tail, which we do not observe. 

Importantly, orbital shear, stellar wind confinement, preferential dayside mass loss, and radiative transfer must all conspire to create the conditions of high enough column density to populate the \ion{He}{1} metastable triplet to detectable levels while imbuing the velocity substructure that we see.  The detailed 3D modeling of the interplay of these effects is beyond the scope of the current work, but is feasible with adaptations to existing 3D simulations \citep{2022ApJ...926..226M}.

Here we interpret the bulk velocity substructure in Figure \ref{fig:centroids} under these conceptual gas dynamics mechanisms.  We attribute the mild pre-transit arc to stellar wind acceleration.  The stellar wind is initially too weak to plow the dense escaping gas until a separation of 60-130 planetary radii ($+0.15$ orbital phase), the inflection point in an enormous bow shock.  Gas at these separations has had enough time to diffuse, both out of the orbital plane and to slightly larger radii shells, in turn lowering the column density along the line-of-sight.  The lower column density provides both fewer Helium atoms to participate in absorption, and less overall NUV shielding, leading to greater fractional ionization and further depopulation of the \ion{He}{1} metastable state.

The weak post-egress tail can be understood as follows.  A weak nightside mass loss means that there is both less overall column density and less NUV photoionization shielding, subduing the overall \ion{He}{1} metastable state's signal strength.  The blueshift arises from the Keplerian shear that lags the planet, and the ever-blueshifting centroid represents the stellar wind's greater ability to carry away the lower total inertia of less nightside material.

In order to examine the in-transit velocity structure, we have to consider the interplay of two subtle geometrical effects.  First, the planet's Roche lobe is small enough (2.7 $R_\mathrm{p}$) that the majority of the projected stellar disk should be filled with escaping material unbound from the planet, even at the time of mid-transit.  In other words, only a small fraction of the \ion{He}{1} excess signal would be expected to trace the planet's motion, as opposed to the hypothetical in-transit signal for WASP-107~b that definitively traces out the planet's orbital path \citep{2022ApJ...926..226M}.

Second, the star's non-negligible rotational velocity sets up a configuration analogous to Doppler Tomography in the Rossiter-McLaughlin (RM) effect, but distinct in a subtle way.  Whereas traditional Doppler Tomography treats the scanning reticle as an opaque planetary disk, here the reticle resembles a transmissive filter with a wavelength center and width that varies in space and time.  The interplay of spatial and spectral illumination and absorption can yield minor second-order effects, and overall we anticipate those effects to be secondary compared to the mere existence of absorbing material spanning the entire stellar disk.

\subsection{Variable planetary wind}
Figure \ref{fig:HPFperCampaign} and \ref{fig:centroids} show differences in the Helium line profiles over months-long and years-long timescales.  The line profiles are qualitatively similar, but show slightly different tendencies towards redshifting.  For example, the centroid of the Helium line during the April 2022 transit appears slightly blueshifted relative to the stellar rest frame.  In comparison, the May 2020 transit probes the same planetary phases, yet resides slightly redshifted at transit midcenter.  We interpret this line profile variability as indicative of genuine planetary wind variability, which can arise from the interplay of stellar and planetary winds \citep{2009ApJ...693...23M} and weather-like feedbacks in the planetary upper atmosphere.

\section{Mechanisms Driving Atmospheric Escape}\label{secPhysMech}

\subsection{XUV Irradiation-driven mass loss}
XUV photoevaporation heats the upper layers in the atmosphere, driving a hydrodynamic wind \citep{2009ApJ...693...23M}.  Photoevaporation stands out as offering a natural cause of the observed leading tail attributable to dayside/nightside differences in mass loss: the greatest supersonic motions arise from the sub-solar point on the planetary dayside.  The effect may be boosted if dayside/nightside energy transport proceeds inefficiently \citep{2009ApJ...693...23M}.  Such a scenario is illustrated in Figure \ref{fig:KeplerianShear}.

The combined effects of photoevaporation, anomalous heating, and tidal gravity are predicted to have especially drastic outcomes for inflated hot Saturns \citep{2023ApJ...945L..36T} such as \hatpb, where the low densities lead to large mass loss rates:

\begin{equation}
    \dot{M} = \frac{3}{4}\frac{\eta F_\mathrm{XUV}}{G K_t \rho_\mathrm{XUV}}\label{thorn23Mdot}.
\end{equation}
 where $K_t$ is the tidal gravity term, $\rho_\mathrm{XUV}$ is the planet density determined using the radius at which XUV radiation is deposited, and $\eta \sim 0.4$ comes from \citet{2022A&A...663A.122C}.  Assuming $\log{L_\mathrm{XUV}/L_\mathrm{bol}}=-4.2$, we would expect \hatpb~ to exhibit $\dot{M}\sim10^2$ M$_\oplus$/Gyr ($2\times10^{13}\;$g/s). At a current $M_\mathrm{p}\sim95\;$M$_\oplus$ and an accelerating mass loss rate, its lifetime would measure in the $<$100 Myr range.  The \citet{2023ApJ...945L..36T} simulations focused on 0.75-1.25 $M_\odot$ host stars; the higher mass $\sim$1.6 $M_\odot$ \hatp would deliver greater tidal gravity for a given semi-major axis, and possibly lower $\log{L_\mathrm{XUV}/L_\mathrm{bol}}$ compared to more active G stars.  Nevertheless, we can project the time series trends in their Figures 3 and 4 to recreate a qualitative history and fate of \hatpb~ under the assumptions of XUV photoevaporation.  Broadly, the planet would have started with an initial mass up to 50$\%$ greater than at present, with $R\sim\;$1.4 $R_\mathrm{Jup}$, for an initial density of $\sim$0.2 g/cm$^{-3}$.  It would lose mass at a rate of tens of Earth masses per Gyr for its 1 Gyr lifetime, expanding modestly until it reaches the critical $\sim$0.1 g/cm$^{-3}$ threshold, at which point the mass loss rate increases to its current value.  It will last only tens of Myr in its current state before losing almost all of its envelope and settling as a $5-15$ M$_\oplus$ core with a final radius of 0.2$-$0.3 R$_\mathrm{Jup}$.

\subsection{Ohmic dissipation-driven mass loss}
The Ohmic dissipation mechanism made two key observable predictions.  Anomalous heating efficiency, $\epsilon$, should initially increase as a function of equilibrium temperature, then degrade as irradiation exceeds equilibrium temperatures of about 2000 K.  Second, hot Saturns ($\lesssim0.5 M_\mathrm{Jup}$) should undergo runaway evaporation, whereas hot Jupiters ($\gtrsim 1 M_\mathrm{Jup}$) should reach stable equilibrium---albeit inflated---radii after Gyr timescales.  Both of these outcomes predated data that could validate them, and these outcomes differ from the behavior of other heating mechanisms, such as photoevaporation or tides.  

Ohmic dissipation may therefore be responsible for both dramatic atmospheric escape and significant radius inflation.  \citet{2011ApJ...738....1B} showed that Ohmic heating acts to inflate hot Jupiters, with planets $\lesssim 0.5 M_\mathrm{Jup}$ overflowing their Roche lobes and leading to evaporation on Gyr timescales.  The Ohmic heating scenario requires only a modest planetary magnetic field ($\gtrsim$1 G) and an equilibrium temperature great enough to thermally ionize some modest fraction of neutral metals, such as the low ionization species of \ion{Na}{1} and \ion{K}{1}.  The ``sweet spot'' for this phenomenon appears to prefer equilibrium temperatures in the range of $1500<T_\mathrm{eq}<2000$ K \citep{2011ApJ...738....1B,2012ApJ...745..138M,2016ApJ...819..116G,2018AJ....155..214T,2022A&A...658L...7K}, where the conductivity is strong enough to cause an effective drag without being so strong that magnetic braking slows the planetary wind.  HAT-P-67~b's $\sim$2000 K sits to the higher end, but still in a region of high Ohmic dissipation heating efficiency.  A proposed order-of-magnitude scaling law predicts an inflation timescale, Eq. 20 in \citet{2011ApJ...738....1B}:

\begin{equation}
    \tau_{\textit{infl}}\sim \left(\frac{0.01}{\epsilon} \right) \left(\frac{M}{M_{J}}\right)^2 \left(\frac{R_{J}}{R}\right)^3 \left(\frac{1500 \textrm{K}}{T_\mathrm{eff}}\right)^4 \textrm{Gyr}. \label{eqInflate}
\end{equation}

\noindent yielding an incredibly short $<$5 Myr timescale for \hatpb, assuming a typical $\epsilon\sim0.01$.  The order of magnitude of this inflation timescale is so fleetingly short that---according to this scenario---\hatpb~must be in the runaway stage of inflation, rapidly losing mass and growing in surface area to fuel a positive feedback loop.  Under this interpretation, \hatpb~ would represent an example of a new category of planet system that is doomed to evaporate entirely due to the Ohmic dissipation mechanism, as predicted by \citet{2011ApJ...738....1B}.  At Roche lobe overflow, a 5 Myr inflation timescale may imply an instantaneous $\dot{M}_\mathrm{infl}>10^{3}$ M$_\oplus$/Gyr.

A few caveats complicate the unambiguous causal interpretation of Ohmic dissipation.  First, the scaling law arguments that produced Equation \ref{eqInflate} were only proposed as coarse estimates, with numerical simulations needed to quantify inflation timescales for individual systems.  Accordingly, a $10\times$ higher inflation timescale of 50 Myr---allowed by the coarse scaling law---would yield $\dot{M}_\mathrm{infl}\sim10^{14}$ g/s, still a very large mass loss rate, and only a factor of 5 away from the baseline 1D model.  Factors of a few uncertainties in the Ohmic dissipation efficiency $\epsilon$ may also contribute.

Second, numerical simulations \citep{2013ApJ...763...13W} and analytic theory \citep{2016ApJ...819..116G} indicate that Ohmic dissipation can stall the contraction of hot Jupiters but cannot easily ``re-inflate'' them after having undergone a traditional cooling curve.  Heat transfer from the atmosphere to a cooled core appears to proceed too slowly, on the order of tens of Gyr \citep{2016ApJ...819..116G}.  This path dependence of Ohmic dissipation would restrict the allowed evolutionary histories of \hatpb, to have arrived at its current location within a few to tens of Myr.  This short time window prefers a physical mechanism such as disk migration, which would be faster than secular eccentric migration with an outer companion. \emph{In-situ} formation of a hot Saturn at these close-in separations may be implausible \citep{2018ARA&A..56..175D}.  

Other caveats like unknowns in planetary magnetic fields, zonal band geometries, and dayside/nightside temperature differences make it impossible to uniquely prescribe Ohmic dissipation, and instead, several additional factors may also be at play \citep{2021A&A...645A..79S}.

\subsection{Reinflation}
Evolved stars increase in luminosity, delivering greater insolation to planets at a fixed separation.  The heightened equilibrium temperature can cause mature planets to inflate, the phenomenon known as \emph{reinflation}.  Such re-inflated hot planets around red giant stars have been recently found by \emph{TESS} \citep{2022AJ....163..120G,2023arXiv230306728G}, though not all planets around evolved stars appear to re-inflate \citep{2022AJ....163...53S}.

\begin{figure}
    \includegraphics[width=\linewidth]{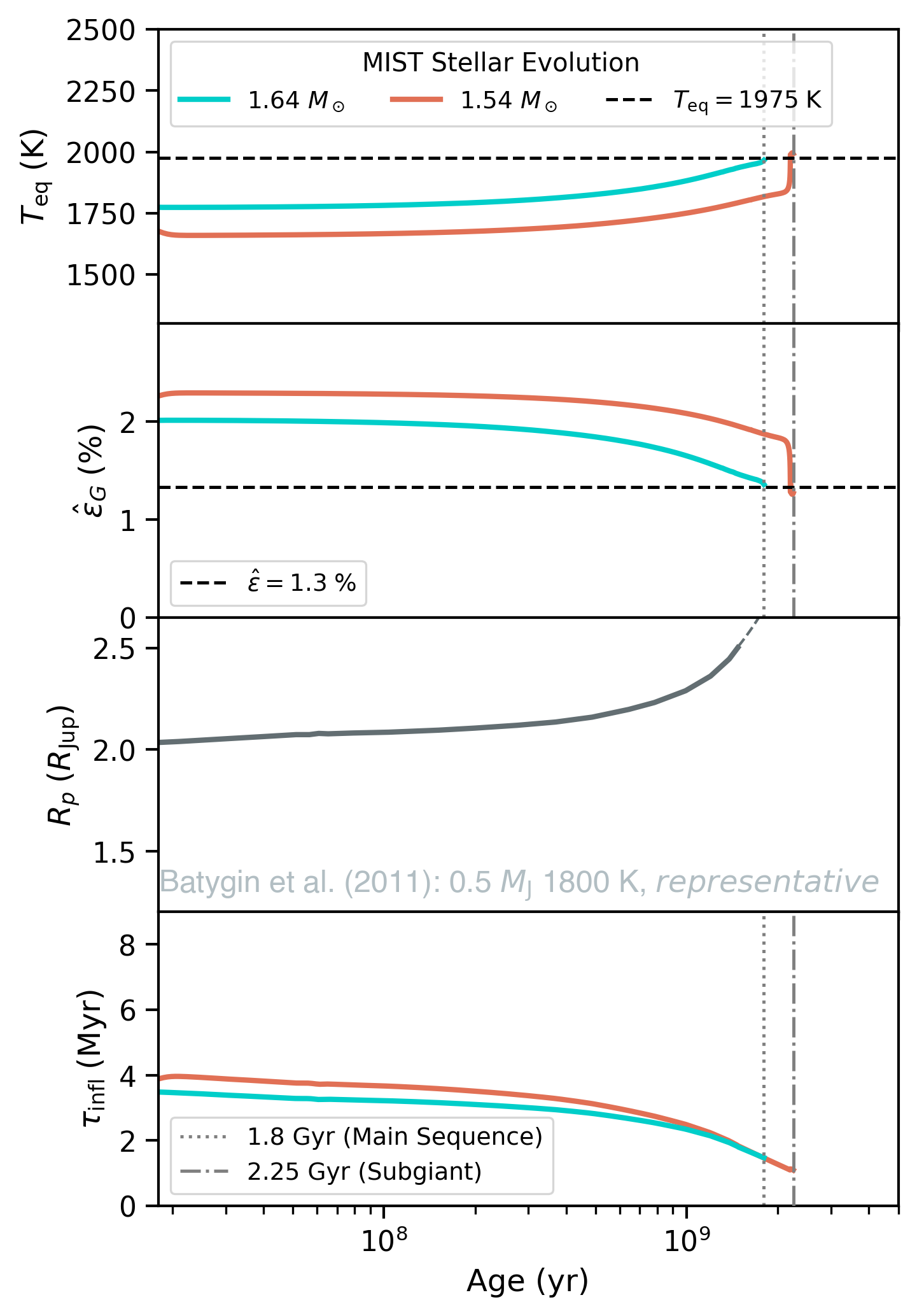}
    \caption{Conceivable evolutionary scenarios for \hatpb.  The planet's equilibrium temperature increases as the stellar luminosity gradually increases on the main sequence.  The second panel from the top shows the corresponding estimate for anomalous heating efficiency from \citet{2018AJ....155..214T}. The Ohmic dissipation model makes predictions for runaway inflation, depending on the incident stellar flux.  The heuristic scaling relation for the inflation timescale predicts vanishingly short inflationary lifetimes in this extreme regime but illustrates the countervailing effects of the top two panels.}
    \label{fig:OhmicInflate}
\end{figure}

\citet{2017AJ....153..211Z} estimated that \hatpb~received about twice the incident flux as a Zero Age Main Sequence (ZAMS) HAT-P-67, based on comparison to the Geneva isochrones. In Section \ref{secMISTtracks} MIST evolutionary tracks showed two equally plausible states: the tail-end of the main sequence or a recently evolved sub-giant.  Figure \ref{fig:OhmicInflate} employs these tracks to quantify the prospects for re-inflation  of \hatpb.  

We further assume that the orbital location has not changed over the system lifetime, that the planet's response to anomalous heating stimulus is instantaneous, and that the anomalous heating efficiency $\hat{\epsilon}_G$ peaks at $T_\mathrm{eq}=1750$ K as derived by \citet{2018AJ....155..214T}.

We conclude that reinflation likely does not have a significant effect on the evolution of HAT-P-67~b: inflation timescales remain relatively unchanged over the planet's lifetime, despite a 50\% spike in incident radiation in the subgiant scenario.  The reason is subtle, as we describe next.  

The second panel from the top illustrates a countervailing effect: the increase in flux actually triggers a decrease in anomalous heating efficiency, $\epsilon$. The greater stellar energy couples into the planet less effectively than the weaker radiation did, roughly balancing out. Together the effects nearly cancel when computing the inflationary lifetimes, yielding nearly identical curves in the bottom panel: a secularly evolving main sequence history gives almost the same inflation timescale as a rapidly increasing subgiant.  It is important to emphasize that the anomalous heating efficiency found by \citet{2018AJ....155..214T} is agnostic to what the actual heating mechanism is: either XUV Irradiation or Ohmic dissipation both have to obey the trend of $\hat{\epsilon}_G(T_\mathrm{eq})$.

The third panel recreates the numerically computed curve for $0.5\;M_\mathrm{Jup}$, $T_\mathrm{eq}=$1800 K from \citep{2011ApJ...738....1B}, which should be considered as representative since it was not necessarily tailored to the evolutionary history of \hatpb.  Nevertheless, the similarity of the curve is remarkable since it arrives at approximately the correct planet radius at the right age for about the right initial mass.

\subsection{Weighing the causes for a lack of sub-Saturns}\label{secLackofSaturns}

\begin{figure*}[t]
    \includegraphics[width=\linewidth]{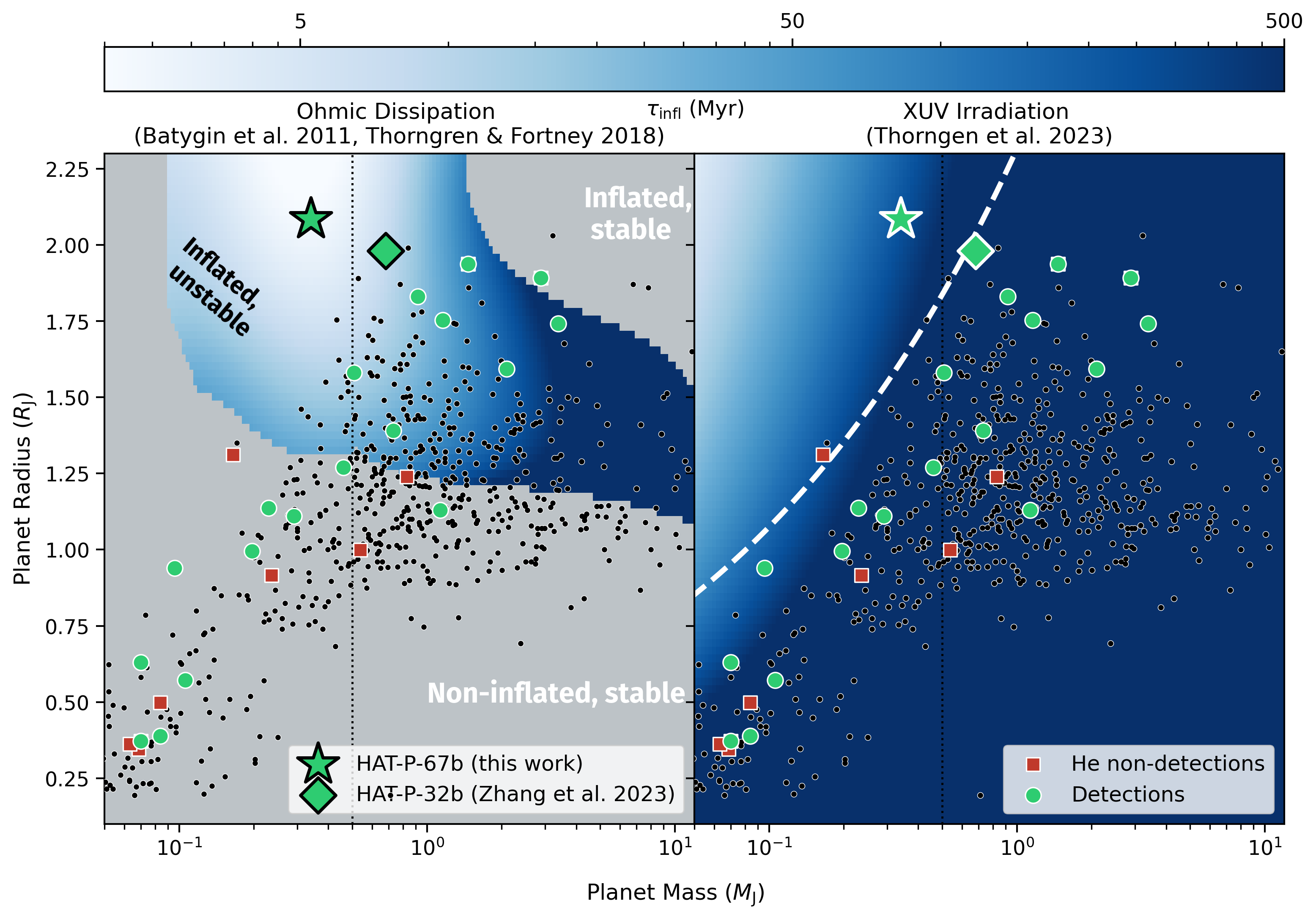}
    \caption{Runaway inflation timescale in the exoplanet mass-radius diagram.  The shading of the left panel shows $\tau_\mathrm{infl}$ from Ohmic dissipation; the right panel shows $\tau_\mathrm{infl}\equiv M/\dot{M}$ from photoionization-driven mass loss.  Individual planets with confident mass and radius detections are shown as small black dots. Helium non-detections are shown in red squares, atmospheric escape detections from any origin in green circles \citep{2022arXiv221116243D}.  Under either scenario, \hatpb~ resides in a sparsely populated region with a short inflationary timescale, making it unstable to runaway evaporation with large mass loss expected.  Photoionization better predicts the depopulation of sources in the upper left low-density region of the diagram, defined by the 0.1 g cm$^{-3}$ iso-density dashed white line.}
    \label{fig:tf2018_theory}
\end{figure*}

When applied to an ensemble of planet systems, both the XUV irradiation and Ohmic dissipation mechanisms expect a void in the planet mass-radius plane. However, the theories make different quantitative predictions for the placement of the dividing lines between stable and unstable populations.  We illustrate these differences in Figure \ref{fig:tf2018_theory}, which shades the mass-radius plane with predictions for the inflation timescale under HAT-P-67b-like conditions.  The XUV irradiation timescale better matches the distribution of planets, with the 0.1g~cm$^{-3}$ density contour setting a conspicuous dividing line in density.  The Ohmic dissipation shading predicts short inflation timescales extending into a region of Jupiter-mass planets with radii commonly observed.  The better match of observed exoplanet demographics to the XUV irradiation shading disfavors Ohmic dissipation.  

Under either scenario, HAT-P-67b resides in an extremely short-lifetime region.  Its nearest analog, HAT-P-32b, offers an interesting test case since it has recently been shown to exhibit a large helium excess \citep{doi:10.1126/sciadv.adf8736}.  It resides just slightly denser than the 0.1 $g\;cm^{-3}$ dividing line, with XUV predicting over 10$\times$ longer inflationary timescale for HAT-P-32b than for HAT-P-67b, while Ohmic dissipation expects merely a factor of 3 difference.  HAT-P-32b exhibits a more symmetric leading and trailing tail, whereas HAT-P-67b stands out as exhibiting evidence for preferential mass loss on the highly irradiated planet dayside.  These differences may make this pair an especially valuable laboratory for developing theories of atmospheric escape.

Finally, we consider the prospect that the planetary mass of HAT-P67b is much lower than the 1-$\sigma$ estimate, such that the white-light planetary radius \emph{is} the Hill radius (we have previously assumed the Hill radius is 2.7 planetary radii).  This extreme scenario could manifest enormous mass loss rates, with the atmosphere's reservoir of mass fleeing the gravitational potential well directly, without the cushion of an exosphere.  This terminal Roche Lobe overflow should have observational consequences.  In particular, heavy elements would easily leak into the planetary wind, yielding possibly many observable metal lines in the UV.

\subsection{Other planets likely to be evaporating}

The key figure-of-merit can be distilled to $\tau_\mathrm{infl}$, which can be thought of as an atmospheric escape spectroscopy metric \citep{2018PASP..130k4401K} for inflated planets in the shaded regime of Figure \ref{fig:tf2018_theory}.  Table \ref{tabHeSM} presents the rank-ordered list of planets by this metric, indicating that they are mostly smaller and more massive than \hatpb.  Only KELT-11~b shows a shorter nominal inflationary timescale than \hatpb, owing to its low mass and equilibrium temperature residing closer to the peak of $\epsilon_G(T_\mathrm{eq})$.  We predict that these few other inflated \hatpb-analogs should show evidence for significant atmospheric escape, comparable to what we see for \hatpb.  These sources make excellent targets for \ion{He}{1} 10833 observations, and/or other atmospheric escape diagnostics.  As emphasized, there are significant uncertainties in the inflation timescale, but at least its order-of-magnitude value gives us a quantitative and justified way to prioritize target selection.

\begin{deluxetable*}{lLLCRR}
    \tablewidth{0pc}
    \tablecaption{
        Coarse Inflation Timescales for Other Systems
        \label{tabHeSM}
    }
    \tablehead{
        \colhead{ }   &
        \colhead{ } &
        \colhead{ }   &
        \colhead{ } &
        \colhead{(OD)} & 
        \colhead{(XUV)} \\
        \colhead{Planet}   &
        \colhead{Mass} &
        \colhead{Radius} &
        \colhead{$T_\mathrm{eq}$} &
        \colhead{$\tau_\mathrm{infl}$} &
        \colhead{$\tau_\mathrm{infl}$} \\
        \colhead{}   &
        \colhead{$M_\mathrm{Jup}$} &
        \colhead{$R_\mathrm{Jup}$}   &
        \colhead{K} &
        \colhead{Myr} & 
        \colhead{Myr}
    }
    \startdata
     HAT-P-67 b &       0.34^{+0.25}_{-0.19} & 2.085^{+0.096}_{-0.071} &   1900^{+25}_{-25} &       4 &          40 \\
  KELT-11 b &              0.171\pm0.015 &              1.35\pm0.1 &   1710^{+51}_{-46} &       3 &          90 \\
 HAT-P-65 b &              0.527\pm0.083 &             1.89\pm0.13 &   1930^{+45}_{-45} &       9 &         100 \\
 WASP-127 b & 0.1647^{+0.0214}_{-0.0172} & 1.311^{+0.025}_{-0.029} &   1400^{+24}_{-24} &       7 &         200 \\
 HAT-P-32 b &        0.68^{+0.11}_{-0.1} &            1.98\pm0.045 &     1840^{+7}_{-7} &      10 &         200 \\
 WASP-153 b &                0.39\pm0.02 &     1.55^{+0.1}_{-0.08} &   1700^{+40}_{-40} &      10 &         300 \\
  HATS-26 b &               0.65\pm0.076 &             1.75\pm0.21 &   1920^{+61}_{-61} &      20 &         400 \\
 Kepler-7 b &    0.441^{+0.043}_{-0.042} &           1.622\pm0.013 &   1630^{+10}_{-10} &      10 &         400 \\
  HATS-56 b &              0.602\pm0.035 & 1.688^{+0.039}_{-0.055} &   1900^{+16}_{-16} &      20 &         400 \\
Kepler-12 b &    0.432^{+0.053}_{-0.051} & 1.754^{+0.031}_{-0.036} &   1480^{+30}_{-30} &      20 &         400 \\
  TOI-954 b &    0.174^{+0.018}_{-0.017} & 0.852^{+0.053}_{-0.062} & 1530^{+123}_{-164} &      20 &         500 \\
 WASP-174 b &               0.33\pm0.091 &            1.437\pm0.05 &   1530^{+17}_{-17} &      10 &         500 \\
 HAT-P-64 b &       0.58^{+0.18}_{-0.13} &            1.703\pm0.07 &   1770^{+22}_{-16} &      20 &         500 \\
 WASP-172 b &                 0.47\pm0.1 &              1.57\pm0.1 &   1740^{+60}_{-60} &      10 &         500 \\
 HAT-P-40 b &                0.48\pm0.13 &             1.52\pm0.17 &   1770^{+33}_{-33} &      20 &         500 \\
    \enddata
    \tablecomments{XUV Inflation timescale assumes $L_X/L_\mathrm{bol}=6.3\times10^{-4}$ (Equation 5 of Sanz-Forcada et al. 2011), which is an overestimate for HAT-P-67 and old/slowly-rotating host stars.}
\end{deluxetable*}

\section{Discussion}\label{secDiscuss}

\subsection{Cause for stellar modulation}
As discussed in Section \ref{TESSmodulation}, the \emph{TESS} lightcurve exhibits up to $0.36\%$ peak-to-valley modulation amplitude, with a characteristic timescale of $P=4.7-5.9$ days. Some sectors exhibit lower amplitudes.  The conventional interpretation would consign these cyclical modulations to the familiar stellar activity: surface features---either starspots, faculae, or plage---entering and exiting the projected stellar disk on the stellar rotation period comparable to the 4.8-day orbital period of planet b.  The cause for such orbital and rotational synchronization would then be either coincidence or secular star-planet tidal interactions.  For the latter, disk migration could have naturally ceased near the co-rotation radius, leaving the planet to reside naturally near the stellar $P_\mathrm{rot}$.  

Alternatively, Star Planet Magnetic Interaction (SPMI) could be at play \citep{2018haex.bookE..25S}.  In this scenario, the magnetic field of the planet permeates the space between the star and the planet.  Magnetic perturbations propagate via planetary magnetic fields with the Alfv\'en speed.  The planet can interact with the star if the Alfv\'en speed exceeds the stellar wind speed controlling the bulk motion of the intervening medium.  The non-detection of variability in the \ion{Ca}{2} H and K lines and H$\alpha$ lines appears to disfavor this SPMI interpretation since SPMI could be expected to cause variations in these diagnostics.

It is hypothetically possible---albeit unlikely---that mass loss could be great enough to produce variability in the wide band \emph{TESS} lightcurve.  The HPF spectra reveal up to 10\% signal depth over a few Angstroms.  The TESS bandpass barely includes 10833 \AA, at a location of diminishing throughput.  The Helium signal alone would manifest as a mere $\sim4$ ppm flux loss when integrated over a TESS-throughput-weighted F-star spectrum--negligible compared to the observed $0.36\%$ peak-to-valley modulation.  An ensemble of additional lines in the red-optical cannot realistically resemble the TESS modulation.  The inventory of such conceivable atomic lines detectable from the planet in the TESS bandpass numbers merely a few, with \ion{Ca}{2} infrared triplet and H$\alpha$ chief among them \citep{2023arXiv230606971L}.   A putative H$\alpha$ line would have to be about 30\% deep over 10 \AA~ wide to manifest perceptibly in the TESS data.  Such an implausibly deep and wide line likely would have been observed as perturbations in the Keck HIRES spectra, even with its limited phasing.  Hypothetically, dust dredged up in the mass loss process could cause a large enough broadband continuum flux loss to manifest in TESS.  We may expect to see some variable reddening in that case.  

\subsection{Implications for exosphere detection in non-transiting planets}

We measured a large extent of Helium escape along the arc of the orbit, but we necessarily can place only coarse constraints on the vertical extent---in the direction perpendicular to the orbital plane, $\pm z$.  We estimate the detectable vertical extent must be at least a few stellar radii, such that we still could have detected Helium escape in a hypothetical HAT-P-67b-like system even if it were non-transiting, in a ``near miss'' configuration.  We propose a new category of semi-transiting planet, dubbed ``exospheric grazers'', in which the planet does not produce a detectable white-light transit depth, but \emph{does} produce measurable \emph{line}-based exospheric absorption.  This category appears to have been neglected due to the assumption that exospheres were only easily detectable at several planetary radii.  While large tails have been seen previously in Ly$\alpha$, the scarcity of UV resources prohibited searches of this kind.  The discovery of large Helium tails in HAT-P-32~b \citep{doi:10.1126/sciadv.adf8736} and now HAT-P-67~b suggest that the identification of these exospheric grazers may be achievable with current instrumentation and possibly existing archival data from near-IR RV-planet searches.  Non-transiting, strongly irradiated planets with well-constrained orbits may make ideal targets for the detection of this phenomenon.

\section{Conclusions}

We have presented a multi-year spectroscopic survey of HAT-P-67~b, a low-density, heavily irradiated Saturn-mass (or lower) planet.  We identified a large leading tail, with a much weaker trailing tail, which we found to be direct evidence of preferential dayside mass loss.  HAT-P-67~b stands out as an outlier in the mass-radius plane, which we examine through the lens of different mechanisms for anomalous heating.  Both XUV irradiation and Ohmic dissipation predict runaway inflation for such inflated hot Saturns, and we quantitatively weigh these two scenarios, finding XUV irradiation as more straightforwardly predictive of the overall demographic population of exoplanets observed to date.  

We report in-transit line profile variability, which we attribute to the delicate interplay of planetary and stellar winds.  We identify several avenues for future work, including additional monitoring of the line profile variability to probe the stellar-and-planetary wind interaction.  The large signal should be perceptible in other spectral tracers, such as metal lines in the UV.  We offer a list of other planets that may be likely to exhibit mass loss under the Ohmic dissipation and XUV irradiation scenarios.

\appendix

\section{Spectral variability non-detections} \label{appendixSec}
\subsection{Calcium IR Triplet and other line diagnostics}
\begin{figure}
    \includegraphics[width=0.5\linewidth]{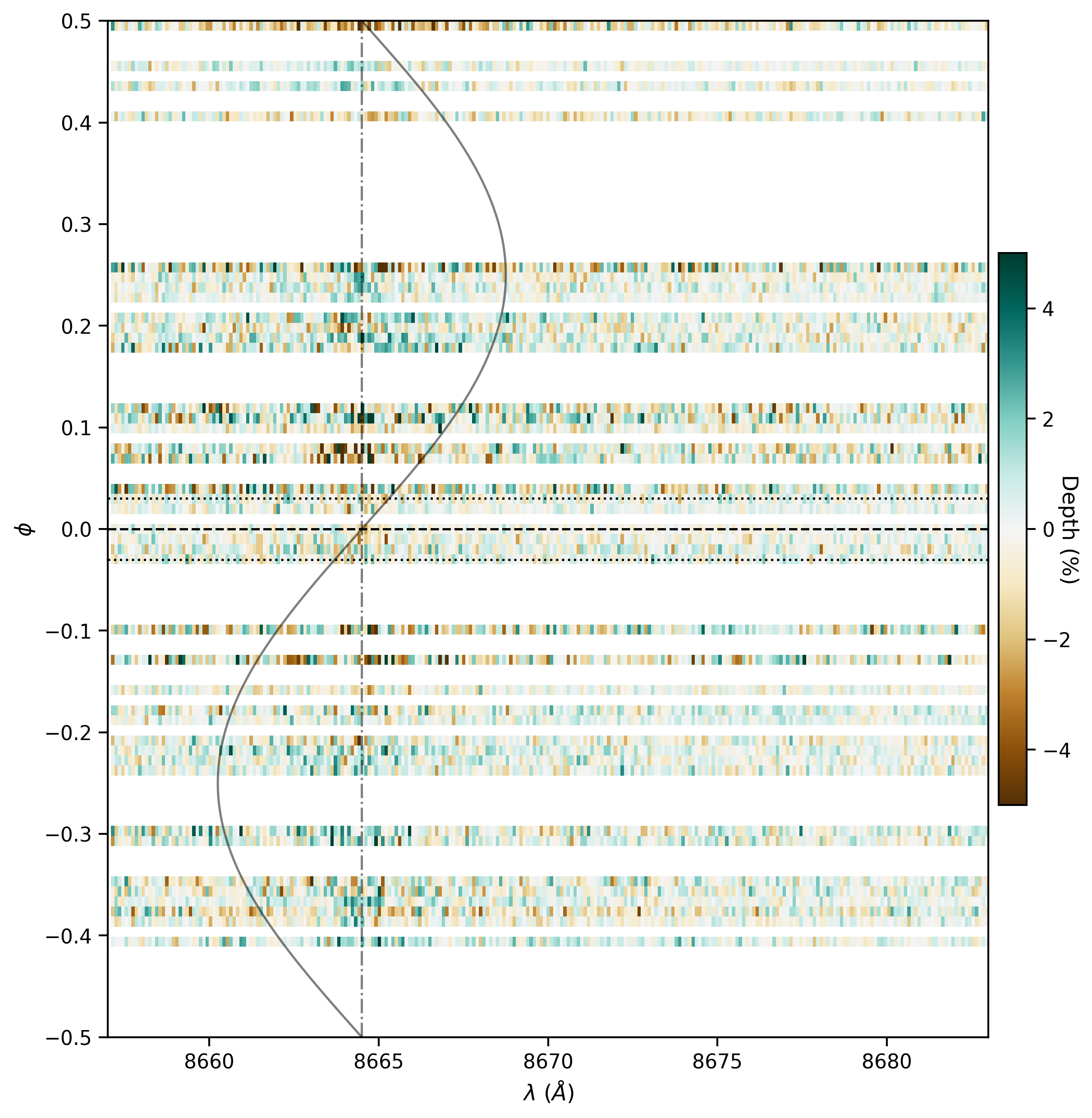}
    \caption{Non-detection of variability in the Calcium IR Triplet line at 8862 \AA.  No other HPF lines appear to show any variability.}
    \label{fig:CaPhaseScan}
\end{figure}

We examined the \ion{Ca}{2} IR Triplet lines (Ca IRT) for evidence of variability.  No variation was seen in these features, which were pre-processed in the same way as the Helium feature.  Figure \ref{fig:CaPhaseScan} shows a similar phase plot as Figure \ref{fig:HPFscanResid}, adapted to the line at 8662 \AA.  We see no conspicuous variability at this or any of the other Ca IRT lines.  We also found no detectable variability in the Pa$\delta$ line at 10050 \AA, nor other deep lines at 10330 \AA~ and elsewhere.  \ion{He}{1} 10833 appears to be the only conspicuously variable line in the HPF spectrum.

\begin{acknowledgements}

    This material is based upon work supported by the National Aeronautics and Space Administration under Grant Number 80NSSC21K0650 for the ADAP program, 80NSSC20K0257 for the XRP program, and 80NSSC22K0181 through the TESS Guest Investigator program issued through the Science Mission Directorate.  

    C.V.M. acknowledges support from the Alfred P. Sloan Foundation under grant number FG-2021-16592.  
    
    Support for program HST-AR-15805.001-A was provided by NASA through a grant from the Space Telescope Science Institute, which is operated by the Associations of Universities for Research in Astronomy, Incorporated, under NASA contract NAS5- 26555.
    
    Based on observations obtained with the Hobby-Eberly Telescope (HET), which is a joint project of the University of Texas at Austin, the Pennsylvania State University, Ludwig-Maximillians-Universitaet Muenchen, and Georg-August Universitaet Goettingen. The HET is named in honor of its principal benefactors, William P. Hobby and Robert E. Eberly.

    These results are based on observations obtained with the Habitable-zone Planet Finder Spectrograph on the HET. The HPF team acknowledges support from NSF grants AST-1006676, AST-1126413, AST-1310885, AST-1517592, AST-1310875, ATI 2009889, ATI-2009982, AST-2108512, AST-2108801, and the NASA Astrobiology Institute (NNA09DA76A) in the pursuit of precision radial velocities in the NIR. The HPF team also acknowledges support from the Heising-Simons Foundation via grant 2017-0494.

    The Center for Exoplanets and Habitable Worlds is supported by the Pennsylvania State University and the Eberly College of Science.

    GS acknowledges support provided by NASA through the NASA Hubble Fellowship grant HST-HF2-51519.001-A awarded by the Space Telescope Science Institute, which is operated by the Association of Universities for Research in Astronomy, Inc., for NASA, under contract NAS5-26555.

    This paper includes data collected with the TESS mission, obtained from the MAST data archive at the Space Telescope Science Institute (STScI). Funding for the TESS mission is provided by the NASA Explorer Program. STScI is operated by the Association of Universities for Research in Astronomy, Inc., under NASA contract NAS 5–26555.

    This research has made use of the Keck Observatory Archive (KOA), which is operated by the W. M. Keck Observatory and the NASA Exoplanet Science Institute (NExScI), under contract with the National Aeronautics and Space Administration.

    This research has made use of the NASA Exoplanet Archive, which is operated by the California Institute of Technology, under contract with the National Aeronautics and Space Administration under the Exoplanet Exploration Program.

    This research has made use of NASA's Astrophysics Data System.
\end{acknowledgements}

\facilities{HET (HPF), TESS, ASAS, Exoplanet Archive}

\software{  \texttt{pandas} \citep{mckinney10},
    \texttt{emcee} \citep{foreman13},
    \texttt{matplotlib} \citep{hunter07},
    \texttt{numpy} \citep{2020NumPy-Array},
    \texttt{scipy} \citep{2020SciPy-NMeth},
    \texttt{ipython} \citep{perez07},
    \texttt{seaborn} \citep{waskom14},
    \texttt{astropy} \citep{2022ApJ...935..167A},
    \texttt{muler} \citep{2022JOSS....7.4302G},
    \texttt{lightkurve}, \citep{geert_barentsen_2019_2565212},
    \texttt{telfit} \citep{2014AJ....148...53G},
    \texttt{exoplanet} \citep{exoplanet:joss},
    \texttt{jupyter} \citep{Kluyver2016jupyter},
    \texttt{p-winds}, \citep{2022A&A...659A..62D},
    \texttt{HxRGproc}, \citep{2018SPIE10709E..2UN}
}

\bibliography{ms}

\end{document}